\documentclass[11pt,a4paper]{article}
\usepackage{graphicx}  \usepackage{mathptmx}
\usepackage{amsmath, amssymb}
\usepackage{gensymb}
\usepackage{enumitem}
\usepackage{tikz}
\usepackage{tikz-cd}
\usepackage{wrapfig}
\usetikzlibrary{decorations.pathmorphing}
\tikzset{snake it/.style={decorate, decoration=snake}}
\usepackage{xpatch}
\topmargin = - 1.5 cm
\makeatletter
\@addtoreset{equation}{section}
\makeatother

\newcommand{\id}[1]{\ensuremath{\mathrm{id}}}

\newcommand{\third}{\mbox{\footnotesize $\frac{1}{3}$}}

\newcommand{\quar}{\mbox{\footnotesize $\frac{1}{4}$}}
\newcommand{\half}{\mbox{\footnotesize $\frac{1}{2}$}}
\newcommand{\hi}[1]{\emph{\textbf{#1}}}

\newcommand{\er}{\eqref}
 
\newcommand{\beq}{\begin{equation}}
\newcommand{\eeq}{\end{equation}} 
\newcommand{\bea}{\begin{eqnarray}}
\newcommand{\eea}{\end{eqnarray}}

\newcommand{\ovl}{\overline}
\newcommand{\ul}{\underline}

 \newcommand{\til}{\tilde}
\newcommand{\raw}{\rightarrow}

\newcommand{\x}{\times}

\newcommand{\gm}{\gamma} \newcommand{\Gm}{\Gamma}
\newcommand{\dl}{\delta} \newcommand{\Dl}{\Delta}
 \newcommand{\varep}{\varepsilon}

\newcommand{\lm}{\lambda} 
\newcommand{\rh}{\rho} \newcommand{\sg}{\sigma}
\newcommand{\Sg}{\Sigma}  
 \newcommand{\phv}{\varphi}
\newcommand{\ch}{\ch}  
 \newcommand{\Om}{\Omega}

\newcommand{\inv}{^{-1}}

\newcommand{\CJ}{{\mathcal J}} \newcommand{\CI}{{\mathcal I}}

 \newcommand{\R}{{\mathbb R}}
 
  \makeatletter
 \newcommand{\enp}{\hfill\mbox{}\qed}
\makeatletter
\def\moverlay{\mathpalette\mov@rlay}
\def\mov@rlay#1#2{\leavevmode\vtop{%
   \baselineskip\z@skip \lineskiplimit-\maxdimen
   \ialign{\hfil$\m@th#1##$\hfil\cr#2\crcr}}}
\newcommand{\charfusion}[3][\mathord]{
    #1{\ifx#1\mathop\vphantom{#2}\fi
        \mathpalette\mov@rlay{#2\cr#3}
      }
    \ifx#1\mathop\expandafter\displaylimits\fi}
\makeatother

\newcommand{\Mi}{\mathbb{M}}
\newcommand{\mghd}{\textsc{mghd}}
\newcommand{\pde}{\textsc{pde}}

\newcommand{\ali}{\begin{align}}
\newcommand{\elin}{\end{align}}

\newcommand{\n}{\nabla}
\newcommand{\XM}{\mathfrak{X}(M)}
\newcommand{\p}{\partial}

\newcommand{\GR}{{\sc gr}}

\newtheorem{theorem}{Theorem}[section]

\newtheorem{proposition}[theorem]{Proposition}

\newtheorem{definition}[theorem]{Definition}

{ \begin{trivlist}%
  \item[\hskip \labelsep {\bfseries #1}]%
}%
{ \end{trivlist}%
}
\newcommand{\qed}{\nobreak\hfill$\Box$}

\begin{document}
\pagenumbering{arabic} \setlength{\unitlength}{1cm}\cleardoublepage
\date\nodate
\begin{center}
\begin{LARGE}
{\bf Singularities, black holes, and cosmic censorship: \smallskip

A tribute to Roger Penrose}\end{LARGE}
\bigskip

\begin{Large}
 Klaas Landsman
 \end{Large}
 \medskip
 
 \begin{large}
  Department of Mathematics, Radboud University, Nijmegen, The Netherlands\\
Email:
\texttt{landsman@math.ru.nl}
\end{large}
\smallskip
 \begin{abstract} 
\noindent 
In the light of his recent (and fully deserved) Nobel Prize, this pedagogical paper draws attention to a fundamental tension that drove Penrose's work on general relativity. His 1965 singularity theorem (for which he got the prize) does not in fact imply the existence of black holes (even if its assumptions are met). Similarly, his versatile definition of a singular space-time does not match the generally accepted definition of a black hole (derived from his concept of null infinity). To overcome this, Penrose launched his cosmic censorship conjecture(s), whose evolution we discuss. In particular, we review both his own (mature) formulation and its later, inequivalent reformulation in the \pde\ literature. As a compromise, one might say that  in ``generic'' or ``physically reasonable'' space-times, weak cosmic censorship postulates the \emph{appearance and stability of event horizons}, whereas  strong cosmic censorship asks for the \emph{instability and ensuing disappearance of Cauchy horizons}.
As an \emph{encore}, an appendix by Erik Curiel  reviews the early history of the \emph{definition} of a black hole.
\end{abstract}\end{center}
\tableofcontents

 \begin{center}
\includegraphics[width=0.5\textwidth]{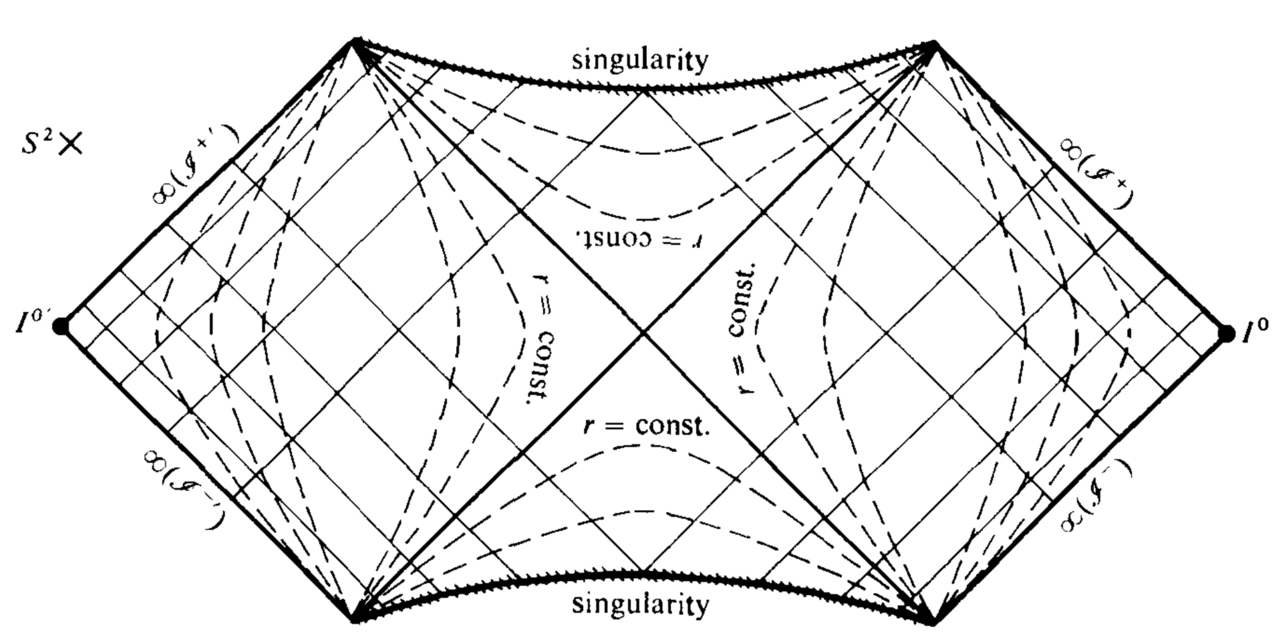} 
\end{center}
\emph{Conformal diagram (Penrose 1968, p.\ 208, Fig.\ 37): `The Kruskal picture with conformal infinity represented.'
Penrose usually drew his own figures in a  professional, yet playful and characteristic way. }

\thispagestyle{empty}
\renewcommand{\thefootnote}{\arabic{footnote}}
\newpage \setcounter{footnote}{0}
\section{Historical introduction}\label{HI}
Roger Penrose got half of the 2020 Physics Nobel Prize `for the discovery that black hole formation is a robust prediction of the general theory of relativity'. This prize was well deserved, since, jointly with Hawking and others, Penrose  has shaped our (mathematical) thinking about general relativity (\GR) and black holes since the 1960s--70s. He would also deserve the Abel Prize for this,  shared with Yvonne Choquet-Bruhat: their combination would  highlight the fact that two originally distinct traditions in the history of mathematical \GR\ have now converged. In the wake of the work of Einstein (1915), these traditions may be said to have originated with Hilbert (1917) and Weyl (1918ab), respectively, as follows. 

 It would be fair to say that Hilbert mainly looked at \GR\ from the point of view of \pde s,\footnote{See 
   Stachel (1992) for an analysis of Hilbert's contribution, as well as for the history of the Cauchy problem of \GR\ up to Choquet-Bruhat, whose contributions were reviewed by  Ringstr\"{o}m (2015) as well as by herself 
    (Choquet-Bruhat, 2014, 2018).} whereas Weyl--once Hilbert's PhD student in functional analysis--had a more geometric view, combined with an emphasis on causal structure.  These different perspectives initially developed  separately, in that the causal theory did not rely on the \pde\ theory whilst for a long time the  \pde\ results were  local in nature.  Penrose contributed decisively to the causal approach to \GR, with its
 characteristic emphasis on the \emph{conformal structure},\footnote{Inspired by special rather than general relativity,  Robb (1914, 1936), Reichenbach (1924),  Zeeman (1964), and  others axiomatized causal structure as a specific \emph{partial order}, as Penrose knew well. } i.e.\ the equivalence class of the metric tensor $g$ under a rescaling $g_{\mu\nu}(x)\mapsto e^{\lm(x)} g_{\mu\nu}(x)$, 
  with $\lm$ an arbitrary smooth function of space and time. Though Weyl (1918b), p.\ 397,  mentions the analogy with Riemann surfaces,\footnote{Riemann surfaces  may equivalently be defined as either one-dimensional complex manifolds or as  two-dimensional Riemannian manifolds \emph{up to conformal equivalence}. 
 Modestly, Weyl does not cite his own decisive contribution to their theory (Weyl, 1913).  This equivalence undoubtedly also influenced Penrose's work on \GR\  and its derivatives like twistor theory. } his real argument for conformal invariance was  that what he calls \emph{Reine Infinitesimalgeometrie} must go beyond Riemannian geometry, which (so he thinks) suffers from the inconsistency that parallel transport of vectors (through the metric or Levi-Civita connection, a concept Weyl himself had co-invented) preserves their length. This makes length of vectors an absolute quantity, which a  `pure infinitesimal geometry' or a theory of \emph{general} relativity should not tolerate. To remedy this, Weyl introduced the idea of \emph{gauge invariance},  in that the laws of nature should be invariant under the above rescaling. To this end, he introduced what we now call a gauge field $\phv=\phv_{\mu}dx^{\mu}$ and
 a compensating transformation $\phv_{\mu}(x)\mapsto  \phv_{\mu}(x)-\p_{\mu} \lm(x)$, and identified $\phv$ with the electromagnetic potential (i.e.\ $A$). Dancing to the music of time, he then proposed that the pair $(g,\phv)$ describes all of physics. The idea of gauge symmetry has lasted and forms one of the keys to modern high-energy physics and quantum field theory: 
   though misplaced in the \emph{classical gravitational} context in which he proposed it,\footnote{See Einstein's negative reaction to Weyl (1918b) in Einstein (2002), Doc.\ 8. See also Goenner (2004), \S 4.1.3. }  through the Standard Model it has ironically become a cornerstone of \emph{non-gravitational quantum} physics! 

The conformal structure of a Lorentzian manifold determines the light cones, and as such
 Weyl was not the only author to discuss causal structure. For example, Einstein (1918) himself wondered if gravitational waves propagate with the speed of light, and showed this in a linear approximation; Weyl mentions this also.\footnote{See e.g.\ page 251 of the English translation of the fourth edition of  \emph{Raum - Zeit - Materie} (Weyl, 1922).} 
The themes of gravitational radiation, conformal invariance,  and causal order were combined and came to a head in the work of Penrose,\footnote{This  history largely remains to be written (Dennis Lemkuhl is working on this). For now, see e.g.\ Thorne (1994), Frauendiener  (2000), Friedrich (2011),  Wright (2013, 2014), and Ellis (2014). Furthermore, both  the written AIP interview by Lightman (1989) and the videotaped interview by Turing's biographer Hodges (2014) are great and intimate portraits of Penrose.
 } who also received additional inspiration from the Dutch artist M.C. Escher and the spinor theory of Dirac.
 In Penrose (1963, 1964, 1965, 1966, 1968, 1972) he introduced most of the global causal techniques and topological ideas   that are now central to any serious mathematical analysis of both \GR\ and Lorentzian geometry (O'Neill, 1983; Minguzzi, 2019). An important exception is global hyperbolicity, which has its roots in the work of Leray (1953) and was adapted to \GR\ by Choquet-Bruhat (1967) and Geroch (1970). Global hyperbolicity is the main concept through which the causal theory meets the \pde\ theory, but Penrose  hardly worked on the \pde\ side.\footnote{Although he was well aware of it:
 the singularity theorem in Penrose (1965) assumes the existence of a Cauchy surface.}

The second piece of history one needs in order to understand Penrose's  contributions that are relevant to his Nobel Prize, is astrophysical. Briefly:\footnote{See  Israel (1987), Luminet (1992), Thorne (1994),  Melia (2009),  Sanders (2014), 
Curiel (2019a), and Falcke (2020)  for history, 
and Misner, Thorne, \& Wheeler (1973), Joshi (1993, 2007),   Poisson (2004), and  Weinberg (2020) for theory.} relatively light stars retire as white dwarfs, in which nuclear burning has ended and inward gravitational pressure is stopped by a degenerate electron gas. In 1931 Chandrasekhar  discovered that this works only for masses $m\leq 1.46 M_{\odot}$, where $M_{\odot}$ is the solar mass. Heavier stars collapse into neutron stars (typically after a supernova explosion), but also these have an upper bound on their mass, as first suggested by Oppenheimer \& Volkoff (1939); the current value is about $2.3  M_{\odot}$.  Stars that are more massive cannot stop their gravitational collapse and unless they get rid of most of their mass/energy they collapse completely.\footnote{Supermassive black holes like Sagittarius A* and M87* are probably formed by mergers and accretion rather than collapse.}
 But what  does this mean mathematically?

 Most of the early intuition came from the Schwarzschild solution, seen as a  model of the final state of such a collapse. This solution is spherically symmetric, and by Birkhoff's theorem any such vacuum solution must be Schwarzschild (or Minkowski). It has
 two very notable features, namely \emph{a curvature singularity as $r\raw 0$} and \emph{an event horizon at $r=2m$}.
Here it should be mentioned that  initially both  caused great confusion, even among the greatest scientists involved such as Einstein and Hilbert, though  Lema\^{i}tre was ahead of his time.\footnote{See Tipler, Clarke, Ellis (1980),  Godart (1992),
 Eisenstaedt (1993), Thorne (1994), Earman (1995, 1999),  and Earman \& Eisenstaedt (1999). Our understanding of the event horizon as a one-way membrane is usually attributed to Finkelstein (1958).}  Apart from their exact locations, these two features, then, may be taken to be the defining characteristics of a black hole, \emph{but especially the event horizon, which is held to be responsible for the ``blackness'' of the ``hole''.} 
 However, even short of a correct technical understanding of these features,
 from the 1920s until the 1950s most leading researchers in \GR\ (including Einstein, Eddington, as well as Landau's school in the Soviet Union, which covered all of theoretical physics) felt that at least the singularity was an artifact of the perfect spherical symmetry of the solution (and likewise for the big bang as described by the spherically symmetric Friedman/{\sc flrw} solution). This negative view also applied to the 
first  generally relativistic collapse model (Oppenheimer \& Snyder, 1939), now seen as groundbreaking, which is spherically  symmetric and therefore terminates in the Schwarzschild solution.

The achievement  usually attributed to Penrose (1965), culminating in his Nobel Prize, is that he 
settled (in the positive) the question whether a more general (i.e.\ non-spherically symmetric) collapse of sufficiently heavy stars (etc.) also leads to a black hole. But
if anyone understood this was \emph{not} the case for the construal of a black hole as an astrophysical object with event horizon,
  it was Penrose himself! He must have been the first to recognize that his singularity theorem from 1965 did not prove the existence of black holes; under suitable hypotheses (involving both the concentration of matter and the causal structure of space-time) it  proved merely the existence (but not even the precise nature or location) of incomplete null geodesics. As such, this implies neither the existence of a \emph{curvature} singularity (not even if the solution is close to Schwarzschild), nor that of an event horizon. 
   Leaving the former aside for the moment, the latter became the topic of what is now called  the  \emph{weak cosmic censorship conjecture}:\footnote{\label{PD} It is worth stressing that Penrose included a genericity restriction right from the beginning, \emph{pace}  Dafermos (2012), p.\ 55.
 The emphasis on initial data in the second formulation does not recur in the strong version of cosmic censorship below, but it is unavoidable in any form of weak cosmic censorship in order to exclude the naked big bang singularity from the conjecture: the point of the second (1979) formulation is that the singularity lies to the future of the Cauchy hypersurface in question. 
    }
  \begin{quote}
\begin{small}
We are thus presented with what is perhaps the most fundamental question of general-relativistic collapse theory, namely: does there exist a ``cosmic censor'' who forbids the appearance of naked singularities, clothing each one in an absolute event horizon? In one sense, a  ``cosmic censor'' can be shown \emph{not} to exist. For it follows from a theorem of Hawking that the ``big bang'' singularity is, in principle, observable. But it is not known whether singularities observable from outside will ever arise in a generic \emph{collapse} which starts off from a perfectly reasonable nonsingular initial state.  (Penrose, 1969,  p.\ 1162)

A system which evolves, according to classical general relativity with reasonable equations of state, from generic non-singular  initial data on a suitable Cauchy hypersurface, does not develop any spacetime singularity which is visible 
from infinity. (Penrose, 1979, p.\ 618)
\end{small}
\end{quote}
Following Penrose (1979) we give a precise mathematical version in \S\ref{Defs}, but in any case it should be clear that in order to prove the existence of black holes from suitable assumptions one needs \emph{both} Penrose's singularity theorem (which gives at least some kind of singularity) \emph{and} the weak cosmic censorship hypothesis (which gives the event horizon): the latter is the missing link between theorem and reality. 

Expanding the scope of cosmic censorship, Penrose (1979), p.\ 619,  subsequently argued that:
  \begin{quote}
\begin{small}
It seems to me to be comparatively unimportant whether the observer himself can escape to infinity. Classical general relativity is a scale-invariant theory, so if locally naked singularities occur on a very tiny scale, they should also, in principle, occur on a very large scale in which a `trapped' observer could have days or even years to ponder upon the implications of the uncertainties introduced by the observations of such a singularity. (\ldots) Indeed, for inhabitants of recollapsing closed universes (as possibly we ourselves are) there is no `infinity', so the question of being locally `trapped' is one of degree rather than principle.  
It would seem, therefore, that if cosmic censorship is a principle of Nature, it should be formulated in such a way as to preclude such \emph{locally} naked singularities.
\end{small}
\end{quote} 
This ban is called \emph{strong cosmic censorship}, which as first shown by Penrose (1979) himself, comes down to the requirement of global hyperbolicity (see \S\ref{CCC}). However: global hyperbolicity \emph{of which space-time?}

More generally, in Penrose's singularity theorem as well as in his two versions of cosmic censorship, 
  it is ambiguous to which space-times the theorem and the hypothesis are applied. Traditionally, in  \GR\ one typically studied
analytically extended solutions to the Einstein equations like--in the context of black holes--the Kruskal extension of the Schwarzschild solution, and similarly (but now not doubled but ``infinitely extended'') for  Reissner--Nordstr\"{o}m and Kerr.
 The  \pde\ approach to \GR, on the other hand,  is based on two slogans, appealing to the fundamental theorem of Choquet-Bruhat \& Geroch (1969):\footnote{A \emph{Cauchy surface} in a space-time $(M,g)$ is a subset $\Sigma\subset M$ such that each endless timelike curve intersects $\Sigma$  exactly once. This makes $\Sigma$ a closed connected $3d$ submanifold of $M$ which can  be chosen space-like and hence Riemannian. 
 A space-time is \emph{globally hyperbolic} iff it has a Cauchy surface. 
 Non-characteristic initial values for the Einstein equations form a triple $(\Sg,h,K)$, i.e.\  a $3d$
 Riemannian manifold  $(\Sg,h)$ equipped with an additional symmetric tensor  $K_{ij}$, satisfying four constraint equations.
 A \emph{Cauchy development} of such initial data is a triple $(M,g,i)$,  where $(M,g)$ is a $4d$ space-time solving the Einstein equations, and $i:\Sigma\raw M$ is an embedding such that $i^*g=h$ and $i(\Sg)$ is a Cauchy surface in $M$ with  extrinsic curvature  $K$. Hence $(M,g)$ is globally hyperbolic.  Choquet-Bruhat \& Geroch (1969) showed that 
 there exists a  Cauchy development of the given $(\Sg,h,K)$, called the
 \emph{maximal  globally hyperbolic development} (\mghd), that is maximal \emph{as a globally hyperbolic space-time solving the  Einstein equations with Cauchy surface $i(\Sg)$ and given  $(h,K)$}. This \mghd\ is unique up to time-orientation-preserving isometries preserving $\Sg$, i.e.\ if $(M,g,i)$ and $(M',g',i')$ qualify then there is an isometry $\psi:M\raw M'$ such that $\psi\circ i=i'$. See Choquet-Bruhat (2009) and Ringstr\"{o}m (2009) for introductions to the \pde\ approach, supplemented by Sbierski (2016). 
 \label{CGBfn}
 }
\begin{itemize}
\item All valid \emph{assumptions} are  about  initial data;
\item All valid \emph{questions} are  about the maximal  globally hyperbolic development (\mghd) thereof. 
\end{itemize}
This clearly affects strong cosmic censorship, in that  asking that a physically reasonable space-time be globally hyperbolic is now deemed empty,
 since the \mghd\ of any inital data automatically has this property.  
For similar reasons also Penrose's version of weak cosmic censorship needs to be reformulated. His singularity theorem does make sense for both traditional solutions and \mghd, but the causes of geodesic incompleteness is quite different is these two cases (except for the Kruskal solution).

Written about and by a mathematician, we start in \S\ref{Defs} with \emph{definitions}. In \S\ref{CCC}
we trace the evolution of Penrose's idea of cosmic censorship, which is illustrated by three black hole examples in \S\ref{examples}. In the last section \S\ref{FSC}, the whole story culminates in Penrose's amazing and influential \emph{final state conjecture}.
The conclusion is that although arguably Penrose did not quite achieve what the Nobel Prize committee says, he developed most of the  techniques, saw the need for  singularity theorems (of which he proved the first) as well as cosmic censorship, and, perhaps most importantly,  showed the way to others.\footnote{\label{fnAB} 
In particular, using the \pde\ approach   Christodoulou (1991, 1999b,  2009) finally established  the formation of  black holes both in spherically symmetric collapse models with (scalar field) matter and in vacuum solutions through focusing of gravitational waves, by proving both causal geodesic incompleteness and the existence of an event horizon. See also follow-ups by Klainerman \& Rodnianski (2012) and  Klainerman, Luk, \& Rodnianski (2014), and reviews by Bieri (2018) and Dafermos (2012). For the incorporation of more realistic matter models see e.g.\ 
Burtscher \& LeFloch  (2014) and  Burtscher (2020).}
Finally, an appendix written by Erik Curiel traces the \emph{definitional} history of the concept of a black hole.
 \section{Definitions}\label{Defs}
In mathematical physics it is essential to start from  \emph{definitions} that are physically relevant, mathematically precise, and workable. Penrose had a remarkable gift for this.\footnote{Penrose was clearly very good at capturing the general spirit of the time in mathematical concepts; this is why his ideas so quickly became mainstream, despite the unfamiliarity of even theoretical physicists at the time with a field like topology (see
Thorne, 1994, Chapter 13, which describes  Penrose's  role in the \GR\ community). But he did so in his own  unique individual way: `It was important for me always, if I wanted to work on a problem, to think I had a different angle on it from other people. Because I wasn't good at following where everybody else went. I wasn't the kind of person who could pick up the prevalent arguments and knowledge of the time. Other people were good at that. They could suck it all out and put it together and make advances. I was the kind of person who'd have some kind of quirky way of looking at something on my own, which I would hide away and work at. So it meant that I had to have some way of looking at a problem that was my own.'  (Lightman, 1989). }
For our purpose, i.e.\ what to make of the citation for his Nobel Prize,
 Penrose contributed at least five great definitions to \GR, namely of:
\begin{itemize}
\item \emph{Null infinity}, in turn implying a definition of an event horizon and hence of a black hole;
\item \emph{Trapped surfaces}, formalizing the condition that gravity is strong enough to focus light-rays;
\item \emph{Singularities in space-time}, which he characterized through incomplete causal geodesics;
\item \emph{Weak cosmic censorship}, stating that space-time singularities are covered by event horizons;
\item \emph{Strong  cosmic censorship}, forbidding even nearby causal contact with  space-time singularities.
\end{itemize}
In this section we explain the first three definitions, leaving the last two for a separate section (\S\ref{CCC}).\footnote{
Unexplained notions may be found in the standard \GR\ textbooks such as Wald (1984) or  Chru\'{s}ciel (2019).
A
\emph{space-time} $(M,g)$ is a $4d$ connected time-orientable Lorentzian manifold with time orientation, i.e.\ the metric has signature $(-+++)$ and one has a way of distinguishing past from future by selecting, at each point in a continuous and consistent way, a \emph{forward} and a \emph{backward} light-cone. This  leads to one of Penrose's most effective notations, namely the relation  $J\subset M\x M$,  where  $(x,y)\in J$, also written as $y\in J^+(x)$ or $x\in J^-(y)$ or $x\leq y$, iff  there exists a future-directed (fd) causal curve from $x$ to $y$. For $A\subset M$ we write $J^{\pm}(A)=\cup_{x\in A}J^{\pm}(x)$.
Similarly, $I\subset M\x M$ is defined by replacing `causal' by `timelike'; one writes $x\ll y$ iff  $(x,y)\in I$.}
\subsection{Null infinity}\label{ni}
Null infinity and the ensuing concept  of a black hole are predicated on the following concept:\footnote{\label{fn14} See originally Penrose (1964), who--in the context of gravitational waves--adds the condition that every null geodesic has two end-points on $\p\til{M}$, defining $(M,g)$ to be \emph{asymptotically simple}. In that case each connected component of $\p\til{M}$ is  diffeomorphic to $\R\x S^2$, as is often the case even more generally (and as such is sometimes included in Definition \ref{defAF}).
 See also Hawking \& Ellis (1973),  \S 6.9, Geroch (1977), Wald (1984), \S 11.1, Penrose \& Rindler (1986), Chapter 9, Stewart (1991), Chapter 3, Frauendiener (2000),  
 Valiente Kroon (2016), and Chru\'{s}ciel (2020), \S 3.1.
 The question how Definition \ref{defAF} relates to asymptotic flatness as defined through conditions on the metric, either in space-time ($4d$) or in the initial value problem ($3d$), is very subtle; smoothness of $(\til{M},\til{g})$ implies detailed fall-off (or `peeling') properties of the Weyl tensor at infinity. See e.g.\ 
  Geroch (1977),  Stewart (1991), Klainerman \& Nicol\`{o} (2003),  Friedrich (2004, 2018),
   Adamo, Newman, \& Kozameh (2012),  Dafermos (2012),  Chru\'{s}ciel \&  Paetz (2015), and Paetz (2014).
 However, for the usual  stationary black hole solutions  and more generally for stationary space-times satisfying standard energy conditions the boundary is smooth (Chru\'{s}ciel et al., 2001). }
\begin{definition}\label{defAF}
\begin{enumerate}
\item 
A \emph{conformal completion} of a  space-time $(M,g)$  is a  space-time $(\til{M},\til{g})$, where  $\til{M}$ is a
manifold with boundary,
 with an embedding  $\iota:M\hookrightarrow \til{M}$ such that $\iota(M)=\mathrm{int}(\til{M}):=\til{M}\backslash \p\til{M}$, and $\iota$  is \emph{conformal} in that $\iota^*\til{g}=(\iota^*\Om^2)g$ for some smooth positive function
  $\Om:\til{M}\raw\R^+$ that satisfies:
    \begin{align}
\Om>0\:\:\: \mathrm{ on }\:\:\:  \iota(M); &&  \Om=0 \:\:\:\mathrm{ on } \:\:\:\p \til{M}; && \til{\n}\Om\neq 0 \:\:\: \mathrm{ on } \:\:\:\p \til{M}.
\label{Om3}
\end{align}
\item  $(M,g)$ is \emph{asymptotically flat at null infinity} if it has a conformal completion $(\til{M},\til{g})$ for which:
\begin{enumerate}
\item 
$\p\til{M}=\CI^+\cup\CI^-$, where  $\CI^{\pm}:=\p\til{M}\cap J^{\pm}(M)$, with $J^{\pm}(M)$ computed in $\til{M}$;
\item  The Ricci tensor of the original metric $g$ is such that $R_{\mu\nu}=O(\Om^3)$ pointwise near $\p\til{M}$.
\end{enumerate}
\end{enumerate}
 \end{definition}
In clause 2 and in what follows we tacitly identify $M$ with  $\iota(M)$. Here are some comments on this clause.
\begin{enumerate}
\item[2(a)]  The boundary  $\CI$ (pronounced, as Penrose  suggests, ``scri'') is called \emph{null infinity}. Its components $\CI^+$ and $\CI^-$ are called
 \emph{future null infinity} and  \emph{past null infinity}, respectively.  The idea is that $\CI^+$ ($\CI^-$) consists of limit points of future (past) directed null curves along which $r\raw\infty$. 
 \item[2(a)]  Asking $O(\Om^3)$  is
 on the safe side (one might ask $O(\Om^{2+\varep})$ for $1/2<\varep\leq 1$), and implies that $\Om^{-2} R_{\mu\nu}$ extends  by continuity from $\iota(M)$ to zero on $\p\til{M}$, as in $R_{\mu\nu}(r)\sim 1/r^3$ as $r\raw\infty$.
The simplest way to satisfy this  is to assume that $(M,g)$ solves the vacuum Einstein equations $R_{\mu\nu}=0$; in the presence of matter one equivalently asks that $T_{\mu\nu}$ be $O(\Om^3)$. 
\end{enumerate}
 A crucial fact, noted (\emph{mutatis mutandis}) 
 without proof in Penrose (1964, 1968), is that:\footnote{If $R_{\mu\nu}=\lm g_{\mu\nu}$, then $\til{g}(\til{\nabla} \Om, \til{\nabla}\Omega)=-\third \lm$ on $\p\til{M}$, so that $\til{\nabla}\Omega$ is timelike and hence  $\p\til{M}$ is spacelike if $\lm>0$,  and \emph{vice versa} if $\lm<0$ (Penrose, 1964, Lecture II; Penrose, 1968, p.\ 181). See  Ashtekar,  Bonga, \& Kesavan (2015) and  Ashtekar \&  Magnon (1984), respectively, for these cases.
 But as the king of null geometry in \GR, Penrose must have taken special pleasure in $\lm=0$!
 }
 \begin{proposition}\label{Omeikonal}
On the boundary $\p\til{M}$ the scaling function $\Om$ satisfies the eikonal equation 
\beq
\til{g}(\til{\nabla} \Om, \til{\nabla}\Omega)=0, \label{eikonaleq}
\eeq
so that $\p\til{M}$ (more precisely: each connected component thereof) is a null hypersurface in $\til{M}$.\footnote{ Short of the very subtle regularity issues discussed in footnote \ref{fn14},
 the boundary $\CI$ is smooth, and points like $i^{\pm}$ and $i^0$, typically included in Penrose diagrams, are \emph{not} part of it. However,  if one is interested in \emph{spatial} infinity (Geroch, 1977; Ashtekar, 1980, 2015) one could extend the definition of a conformal completion so as to include these points. } 
\end{proposition}
\emph{Proof}. A simple computation, based on a conformal rescaling of the Ricci tensor,\footnote{It is easily verified by direct computation, and found in many books (Valiente Kroon, 2016, \S 5.2.2; Chru\'{s}ciel, 2020, Appendix H.6)
 that if $g'=\phv^2 g$, then $R'_{\mu\nu}=R_{\mu\nu}-\phv\inv(2\n_{\mu}\n_{\nu}\phv +g_{\mu\nu}\Delta_g \phv)+\phv^{-2}(4\n_{\mu}\phv \n_{\nu}\phv-
g_{\mu\nu}g(\n\phv,\n\phv))$.
 Now replace $g'\leadsto g$ and $g\leadsto\til{g}$, so that $\phv=1/\Om$. This gives 
$R_{\mu\nu}=\til{R}_{\mu\nu}+\Om\inv(2\til{\n}_{\mu}\til{\n}_{\nu}\Om+ \til{g}_{\mu\nu}\Dl_{\til{g}}\,\Om)-3\Om^{-2}\til{g}(\til{\nabla} \Om, \til{\nabla}\Omega)\til{g}_{\mu\nu}$, where  $\til{\nabla} = \til{g}^{\rh\sg}\til{\nabla}_{\rh}\til{\nabla}_{\sg}$. This is eq.\ (11.1.16) in Wald (1984), which immediately yields \er{preeikonal}.} shows that
\begin{equation}
\til{g}(\til{\nabla} \Om, \til{\nabla}\Omega)=\mbox{\footnotesize $\frac{1}{12}$}(\Om^2\til{R}-R)+\half \Om\,\Dl_{\til{g}}\,\Om.\label{preeikonal} 
\end{equation}
Since $\til{g}$ is regular on $\p\til{M}$ (where $\Om=0$) and $R_{\mu\nu}=O(\Om^3)$ gives $R=O(\Om)$, eq.\ \er{eikonaleq}  follows.
 \enp
 
\smallskip 
\noindent 
Following Hawking and Penrose--we leave the tangled history to the appendix--we may then define 
   \begin{align}
  \mathcal{B}:= M\backslash J^-(\CI^+); &&   \mathcal{W}:= M\backslash J^+(\CI^-), \label{BHRWHR}
  \end{align}
  called the \emph{black hole region} and the \emph{white hole region}  in $M$, respectively;
   each connected component of $  \mathcal{B}$, if not empty, is then simply a \emph{black hole}.\footnote{
  See the appendix by Erik Curiel for historical information on this definition.
  See also Thorne (1994), Chapter 7.} It can be shown that $J^{\mp}(\CI^{\pm})$ is open.\footnote{
  Since $I^{\mp}(\CI^{\pm})\cap M=J^{\mp}(\CI^{\pm})\cap M$, one could have used $I$ instead of $J$ in \er{BHRWHR}; see Wald (1984), p.\ 308.}
 The 
boundaries
    \begin{align}
 \mathcal{H}_E^+:=\p \mathcal{B}; &&   \mathcal{H}_E^-:=\p  \mathcal{W}, \label{defEHplus}
  \end{align}
  decompose into the \emph{future} and \emph{past event horizons} of each of the black and white holes in $M$ respectively. 
  Since the hole regions $\mathcal{B}$ and $\mathcal{W}$  are closed, the event horizons form part of  the black/white holes. 
    
The analysis of such space-times is greatly facilitated by Penrose's \emph{conformal diagrams},\footnote{
 These confirm what Penrose often says, namely that he prefers to think in terms of pictures.
Since Penrose started  in algebraic geometry as a PhD student of Hodge in Cambridge, he was undoubtedly influenced by the theory of Riemann surfaces in finding this concept (like Weyl, as mentioned in the historical introduction). For example, \emph{mutatis mutandis} the closed Poincar\'{e} disk is a Penrose diagram of
the Poincar\'{e} upper half plane. Penrose must also have been influenced by the famous \emph{Circle Limit} woodcuts by Escher (nos.\ I--IV, dating from 1958--1960). See also Wright (2013, 2014). 
} now  called  
\emph{Penrose diagrams}.\footnote{Or, sometimes,  \emph{Penrose--Carter diagrams}. Carter himself speaks of \emph{PC-diagrams}, perhaps tongue-in-cheek saying that \emph{PC} stands for \emph{Projective Conformal}. See  Chru\'{s}ciel (2020), Chapter 6 for an axiomatic theory of such diagrams. On a pragmatic case-by-case basis, construct and draw  $\til{M}$,
 suppress two-spheres, and use coordinates in which null geodesics are at $\pm 45\degree$, as in Minkowski space-time (indeed the  $\pm 45\degree$ idea goes back to Minkowski himself, who also drew his own diagrams). 
}
These became an important tool for visualizing black holes (Carter, 1973; Hawking \& Ellis, 1973). The title page shows one of the first such diagrams, drawn by Penrose himself.
 \newpage
 However, the above definition of a black hole, though mathematically sweet, is not uncontroversial:
 \begin{quote}
\begin{small}
This definition depends on the whole future behaviour of the solution; given the partial Cauchy surface $\mathcal{S}(\tau)$,\footnote{A \emph{partial Cauchy surface} $\Sg$ is an acausal edgeless subset of $M$ (Hawking \& Ellis, 1973, p.\ 204; Minguzzi, 2019, p.\ 95). This makes $\Sg$ a closed hypersurface in $M$ which, 
because it is edgeless, is inextendible as an acausal set (though not necessarily maximal in $M$ as such). A  sufficient condition for the existence of a partial Cauchy surface is the existence of a time function; see Minguzzi (2019), Theorems 3.39 and 4.100.  In the \pde\ approach it arises when strong cosmic censorship fails,
 see \S\ref{PDECCC}, and a Cauchy surface 
for the \mghd\ turns into a partial one for the extension. 
In that case $\Sg$ acquires a non-empty \emph{Cauchy horizon} $H_C(\Sg)=\p D(\Sg)$, where $D(\Sg)$ is the domain of dependence of $\Sg$, splitting into past and future ones $H_C(\Sg)=H_C^-(\Sg)\cup H_C^+(\Sg)$. 
 }
  one cannot find where the event horizon is without solving the Cauchy problem for the whole future development of the surface.' (Hawing \& Ellis, 1973, p.\ 319)
  
  [The future event horizon] is the boundary of an interior spacetime region from which causal signals can never be sent to the asymptotic observers, \emph{no matter how long they are prepared to wait.}  The region is therefore ``black'' in an absolute sense.' (Ashtekar \& Galloway, 2005, p. 2)

The idea that nothing can escape the interior of a black hole once it enters makes implicit reference to all future time--the thing can never escape no matter how long it tries. Thus, in order to know the location of the event horizon in spacetime, one must know the entire structure of the spacetime, from start to finish, so to speak, and all the way out to infinity. As a consequence, no local measurements one can make can ever determine the location of an event horizon. That feature is already objectionable to many physicists on philosophical grounds: one cannot operationalize an event horizon in any standard sense of the term. Another disturbing property of the event horizon, arising from its global nature, is that it is prescient. Where I locate the horizon today depends on what I throw in it tomorrow--which future-directed possible paths of particles and light rays can escape to infinity starting today depends on where the horizon will be tomorrow, and so that information must already be accounted for today. Physicists find this feature even more troubling. (Curiel, 2019b, p.\ 29)
\end{small}
\end{quote}
It is amusing how differently even top \GR\ experts (and textbook authors!) respond to this charge:
 \begin{quote}
\begin{small}
[The above definition of an event horizon] is probably very useless, because it assumes we can compute the future of real black holes, and we cannot. (Rovelli, quoted in Curiel, 2019b, p.\ 30)

I have no idea why there should be any controversy of any kind about the definition of a black hole. There is a precise, clear definition in the context of asymptotically flat spacetimes (\ldots) I don't see this as any different than what occurs everywhere else in physics, where one can give precise definitions for idealized cases but these are not achievable/measurable in the real world. (Wald, \emph{ibid.}, p.\ 32)
\end{small}
\end{quote}
What seems at stake here is what may be called  
 \emph{Earman's Principle}:
  \begin{quote}
\begin{small}
While idealizations are useful and, perhaps, even essential to progress in physics, a sound principle of interpretation would seem to be that no effect  can be counted as  a genuine physical effect if it disappears
when the idealizations are removed. (Earman, 2004, p.\ 191)
\end{small}
\end{quote}
Note that two kinds of idealizations are involved in the case of event horizons of black holes:
\begin{enumerate}
\item The ability to know an entire space-time $(M,g)$, either from initial data or by direct construction;
\item The  construction of null \emph{infinity} in terms of which  black holes and  event horizons are defined.
\end{enumerate}
Rovelli's comment seems to apply to the first point but Wald's to the second, in which case  they would not  contradict each other. The need to idealize the idea (!) that an event horizon prevents sending signals from the singularity to observers ``far away'' by taking the latter to mean ``at (null) infinity'' arises because \emph{in general} any \emph{finite} distance could potentially lie within the event horizon.
For \emph{specific space-times} like Kruskal or Kerr, the horizons $\mathcal{H}_E^{\pm}$ as defined in \er{defEHplus} can be explicitly located in $M$ without reference to (null) infinity.\footnote{ 
This is different from the idealization in phase transitions and spontaneous symmetry breaking, where even in exactly solvable models one needs the idealization of the thermodynamic limit to have such effects, at least according to their official  definition. See Butterfield (2011) and Landsman (2017), Chapter 10,  for the way to deal with Earman's principle in these cases.} Even if the space-time is not known explicitly, the event horizon (if it has one) by definition lies at some \emph{finite} distance from the singularity (if it has one). Hence in locating the horizon, the phrase ``at infinity'' could be replaced by ``sufficiently far away'' (from the singularity), which agrees with Earman's principle--and therefore Wald's stance seems valid provided it concerns the second point. 

On the other hand, the second point is predicated on the first, which remains unresolved. Thus we are entering an almost axiomatic approach to \GR\ here, liable to the  famous charge that it has `the advantages of theft over honest toil' (Russell, 1920, p.\ 71).
However, nothing is wrong with an axiomatic approach as long as one can find realistic models for the axioms  (or definitions) that show that they are reasonable. This 
 is the case in Penrose's approach. The fact that we cannot `compute the future of black holes' does not disqualify  the event horizon  as an object of nature we can prove theorems about (whose desirability may be different for theoretical and mathematical physicists). 
 What \emph{is} worrying is the precise relationship between the black hole ``shadow'' in the  
  EHT image of  M87* and the  event horizon as defined by \er{defEHplus}, which we cannot possibly know \emph{now}.\footnote{More precisely, where it was 53.5 million years ago. An additional complication is that  (ignoring the rotation of M87* for simplicity) the edge of the disk is not the event horizon at $r=2m$ but the photon sphere 
 at $r=3m$, further dislocated by optical effects so that we actually see it at $r=\sqrt{27}m$, cf.\ Chru\'{s}ciel  (2019), \S 3.9.6 and  Event Horizon Telescope Collaboration (2019).
 }  This raises epistemological questions about the role of theory in observation, which will not even be addressed here, let alone answered. See also Franklin (2017). 
\subsection{Trapped surfaces}
In their excellent review of Penrose's 1965 singularity theorem, Senovilla \& Garfinkle (2015) explain that all singularity theorems in \GR\ share the following three assumptions (we quote \emph{verbatim}):
\begin{enumerate}
\item[(i)] a condition on the curvature;
 \item[(ii)] a causality condition;
 \item[(iii)] an appropriate initial and/or boundary condition.
\end{enumerate}
In Penrose (1965) condition (i) states that $R_{\mu\nu}\dot{\gm}^{\mu}\dot{\gm}^{\nu}\geq 0$ along all null geodesics $\gm$.
Condition (ii) states that the space-time be globally hyperbolic with \emph{non-compact} Cauchy surface; the topological assumption reflects the idea  that the theorem is supposed to apply to black holes and hence to asymptotically flat space-times. His condition (iii) is the existence of a closed \emph{trapped surface}, which is one of the most important concepts in all of black hole (mathematical) physics.\footnote{See initially Hawking \& Ellis (1973), Chapter 9. The  study of trapped surface formation from the \pde\ point of view  began with
 Schoen \& Yau (1983), who gave initial values that \emph{already contain} trapped surface; see also Alaee, Lesourd, \& Yau (2019). 
Christodoulou (1991, 1999a, 2009) first proved the evolution of asymptotically flat initial data \emph{into} trapped surfaces.  
Later literature may be traced back from  Li \& Yu (2015) and  Athanasiou \& Lesourd (2020). 
See also  references in footnote \ref{fnAB}.}  Here is Penrose's  own definition:
 \begin{quote}
\begin{small}
A \emph{trapped} surface [is] defined generally as a closed, spacelike two-surface $T^2$ with the property that the two systems of null geodesics which meet $T^2$ orthogonally \ul{converge} locally in future directions at $T^2$.
(Penrose, 1965, p.\ 58)
\end{small}
\end{quote}
In the presence of a radial coordinate $r$  as in the Schwarzschild, Reissner--Nordstr\"{o}m, and Kerr solutions, this  condition is equivalent to the (metric) gradient $\nabla r$ being \emph{timelike}, which in the Schwarzschild solution
happens for $r<2m$, and which in the other two (subcritical) cases is the case at least for a while after crossing the  event horizon.
In general, the convergence condition can be stated  in terms of the null hypersurface $C$ generated by the future directed null congruence emanating from some (instantaneous) spacelike two-sphere $S^2$, so that $\p C=S^2$. In terms of a tetrad $(e_1,e_2, L, \ul{L})$
with  $(e_1,e_2)$ spacelike and tangent to $C$ , $L$ null and tangent as well as orthogonal to $C$, and $\ul{L}$ null and pointing off $C$, normalized such that $g(e_i,e_j)= \dl_{ij}$, $g(e_i,L)=g(e_i,\ul{L})=0$ for $i,j=1,2$, and finally
$g(L,\ul{L})=-1$ and of course $g(L,L)=g(\ul{L},\ul{L})=0$, 
 all defined on $C$,  null extrinsic curvatures are $2\x 2$ matrices  $k_{ij}=g(\n_jL(t),e_i(t))$ and  $\ul{k}_{ij}=g(\n_j\ul{L},e_i)$, with traces $\theta=\mathrm{tr}(k)$ and  $\ul{\theta}=\mathrm{tr}(\ul{k})$. 
 Then $S^2$ is trapped iff $\theta<0$ and $\ul{\theta}<0$ throughout $S^2$. 
  This condition is local and there are none of the problems afflicting null infinity (cf.\ \S\ref{ni}). 
\subsection{Singularities in space-time}\label{sst}
Ironically, although this is also seen as one of Penrose's most important contributions to \GR\ (and was immediately recognized as such by his contemporaries like Hawking), his definition of a singular space-time (i.e.\ as being causally geodesically incomplete) has to be inferred from his proof by contradiction of his singularity theorem in Penrose (1965, 1968), to the effect that properties (i), (ii), and (iii) of the previous subsection exclude the possibility that $(M,g)$ is also future null geodesically complete.\footnote{\label{gdf} A geodesic $\gm$, which we take by definition to be affinely parametrized,  is called \emph{complete} if it can be extended to  arbitrary values of its parameter, i.e.\ is defined as a map $\gm:\R\raw M$. It is  \emph{future complete} if it is defined as a map $\gm:[a,\infty)\raw M$ for some $a\in\R$, etc. In the Riemannian case, by the Hopf--Rinow theorem geodesic completeness is equivalent to completeness in the topological metric $d$ derived from the Riemannian metric $g$  as the infimum over the path length (computed from $g$) of all curves connecting two given points. Since a Lorentzian metric no longer defines a topological metric, this result is lost. } 

 Hawking (1966), \S 6.1, much more explicitly discusses the need for a good definition of a singular space-time, upon which he arrives at the contrapositive: `We shall say that $M$ is singularity-free if and only if it is timelike and null geodesically complete.'\footnote{As in Hawking \& Ellis (1973), \S 8.1, Penrose is not mentioned here but there is generic acknowledgement in the Preface.} However, unlike Penrose (1965, 1968), Hawking, and later  Hawking \& Ellis (1973), \S 8.1,
emphatically apply this definition to the case where $(M,g)$ is metrically inextendible, in that it cannot be isometrically embedded as an open submanifold of a larger space-time $(M',g')$, subject to certain regularity conditions on both the manifold and the metric. Adapting a definition given by Clarke (1993), p.\ 10 (who however uses arbitrary curves) we may formalize this by:
\begin{definition}
A space-time is \emph{singular} if it contains an incomplete causal geodesic $\gm:[0,a)\raw M$ such that there is no extension $\theta:M\raw M'$ for which $\theta\circ\gm$ is extendible.
\end{definition}
This refinement of Penrose's  definition was originally proposed in order to avoid trivial cases: removing any point from Minkowski space-time makes it geodesically incomplete, but also think of Schwarzschild for $r>2m$ only. But with hindsight, we can say it makes a big difference to impose inextendibility also in nontrivial cases where strong cosmic censorship fails, as will be explained in due course. 
We therefore follow Penrose in defining a space-time to be singular iff it is causally geodesically incomplete, leaving it open whether it can be extended--indeed his 1965 singularity theorem (or any later version thereof) gives no information about  metric inextendibility at all. 
As we shall see, if the three conditions in Penrose's singularity theorem hold,
 the cases where the space-time in question can or cannot be extended are quite different in so far as the 
  nature of the incompleteness is concerned, and both cases are equally interesting. 
Even apart from this, Penrose's definition is once again  controversial; it ended a long period of confusion, but it did so at a price, as was recognized right from the start. As Geroch (1968),\footnote{Further to this classical paper on singularities,
see also Earman (1995, 1996), Senovilla (1997), and Curiel (1999, 2019a). }
 p.\ 526, states: \begin{quote}
\begin{small}
\begin{enumerate}
\item[(a)] there is no widely accepted definition of a singularity in general relativity;
\item[(b)] each of the proposed definitions is subject to some inadequacy.
\end{enumerate}
\end{small}
\end{quote}
For example, the link between singularities and diverging curvature is lost, although this was the original intuition  from both the Schwarzschild and the Friedman ``singularities''. Furthermore, even within the confines of defining singularities through incomplete curves, singling out (causal) geodesics excludes some interesting space-times intuitively felt to be singular--but this can only be detected through the incompleteness of more general curves (Geroch, 1968, appendix). In fact, Penrose (1979)  did incorporate these at a later stage, as will be discussed in \S\ref{PCCC}. 
But ultimately we side with 
  Hawking and Ellis:\footnote{Reminiscent of the great slogan `\emph{A good definition should be the hypothesis of a theorem}' (attributed to J. Glimm).}
 \begin{quote}\begin{small}
`Timelike geodesic completeness has an immediate physical significance in that it presents the possibility that there could be freely moving observers or particles whose histories did not exist after (or before) a finite interval of proper time. This would appear to be an even more objectionable feature than infinite curvature and so it seems appropriate to regard such a space as singular. (\ldots)  The advantage of taking timelike and/or null incompleteness as being indicative of the presence of a singularity is [also] that on this basis one can establish a number of theorems about their occurrence.' (Hawking \& Ellis, 1973, p.\ 258).
 \end{small}
\end{quote}
\section{Cosmic censorship}\label{CCC}
In this section we review Penrose's original versions of cosmic censorship, followed by \pde\ reformulations now in  use.
Penrose (1979) gave a precise statement of strong cosmic censorship that seems almost forgotten, but  translated this into an equivalent characterization in terms of global hyperbolicity that became very influential. 
Since only seasoned relativists will be able to relate global hyperbolicity to the original ideas behind cosmic censorship  (as reviewed in the historical introduction), we first 
 give a unified formulation of both weak and strong cosmic censorship along the lines of Penrose (1979). 
 \subsection{Cosmic censorship \`{a} la Penrose}\label{PCCC}
 Remarkably, where Penrose (1965) defined singularities in terms of  \emph{incomplete causal geodesics}, 
  Penrose (1979) switches to  \emph{endless timelike curves}. It turns out that the change from `causal' to `timelike' does not matter,\footnote{In the light of the analysis below, this follows from Theorem (2.3) in  Geroch, Kronheimer, \&  Penrose (1972).} but the change from \emph{geodesics} to   \emph{curves} is quite substantial.\footnote{Penrose's timelike curves  are smooth  by convention (Penrose, 1972, pp.\ 2--3). Following Minguzzi (2019),  we prefer to work with \emph{continuous causal} curves, which behave better under limits (e.g.\ smooth timelike curves typically converge 
  uniformly, if they do, to continuous causal curves, whereas limits of the latter, if they exist, lie in the same class). 
 We say that a continuous curve $c:I\raw M$ is \emph{causal} if every point $x=c(t)$ on the curve ($t\in I$) has a normal neigbourhood  $U_x$  such that the unique geodesic  connecting $x$ with
 any later point $y\in U_x$ (with $y=c(t')$ for $t'>t$)  is causal. To analyse such curves we introduce an auxiliary (complete)  Riemannian metric $h$ on $M$ (which always exists), 
with associated topological metric $d_h$ defined as in footnote \ref{gdf}, and defining things like absolute continuity etc.
 A continuous curve $c:I\raw M$ is causal iff (possibly after reparametrization) it is absolutely continuous and a.e.\ differentiable on $I$ with $\dot{c}$ causal. Moreover,  for $[s,u]\in I$ the Riemannian length
$L_h(c_{|[s,u]})=\int_s^udt\, \sqrt{h(\dot{c}_n(t), \dot{c}_n(t))}$ is well defined and finite.  See e.g.\ Theorem 2.3.2 in  Chru\'{s}ciel (2011), \S 2.3, and Theorem A.1 in Candela \emph{et al} (2010).
Since the function $u\mapsto L_h(c_{|[s,u]})$ is strictly increasing and hence invertible, 
any continuous causal curve $c$ may  be parametrized by $h$-arc length.  If an fd (i.e.\ future-directed) continuous causal curve $c:[a,b)$ is parametrized by (or proportional to) $h$-arc length, then $b=\infty$ iff $c$ is future endless (Minguzzi, 2008, Lemmas 2.6 and 2.17). \label{hfn}} 
  Forbidding signaling by singularities thus defined turns out to be equivalent to global hyperbolicity (of all of space-time in case of strong cosmic censorship and of $J\inv(\CI^+)$ in the weak version), which is very neat and may justify this change. However, had the original definition in terms of causal \emph{geodesics}  been used, then presumably some weaker causality condition than global hyperbolicity would have been found.\footnote{It is the second (`converse') part of the proof of Theorem \ref{P79theorem} below that does not work for causal  geodesics instead of  curves, since the curve $\gm$ constructed there is not necessarily a geodesic. This goes back to the definition of domains of dependence and  Cauchy surfaces in terms of causal curves rather than geodesics, and may explain Penrose's (1979) choices. 
 }
 
 In any case, the basic problem is to express mathematically what it means for a signal to emanate from a singularity, since the latter is not part of space-time. Happily, it is precisely his own definition of singularities in terms of incomplete causal 
 geodesics--now general causal curves--that enabled Penrose to overcome this problem, drawing on earlier work (Geroch, Penrose, \& Kronheimer, 1972), as follows.  
 
 An endless causal curve may either be complete, i.e.\ have infinite  length, or incomplete (finite length). In the first case it may either go off to infinity, or hover around in a compact set, which  is impossible in a strongly causal space-time; hence Penrose makes this assumption.  In the second case (also assuming $\gm$ is future directed for simplicity) it may either be thought of as  crashing into a singularity, or leading to the edge of an extendible space-time. If $\gm$ is not endless but has a future endpoint $y$, then
 \beq
 I^-(\gm)=I^-(y). \label{IgmIy}
 \eeq
  If $(M,g)$ is strongly causal, then 
$I^{\pm}(x)=I^{\pm}(y)$ iff $x=y$. The idea, then, is that an \emph{endless} (continuous) causal curve $\gm$ corresponds to an \emph{ideal} point 
 $y$ of space-time, which is not contained in $M$ but is still defined by $I^-(\gm)$, this time without \er{IgmIy}.
 By the above case distinction, at least in strongly causal space-times ideal points may be either points at infinity, or singularities, or  boundary points, respectively.\footnote{\label{GKP72}    Geroch, Kronheimer, \&  Penrose (1972) and in their wake Hawking \& Ellis (1973), \S 6.8, show that
 (assuming strong causality) both real points and ideal points of $M$ correspond to subsets  $U\subset M$ that are:
 (i) open, (ii), past sets, i.e.\ $I^-(U)\subset U$, and (iii) indecomposable, in that $U\neq U_1\cup U_2$ where $U_1$ and $U_2$ have properties (i) and (ii) and are neither empty nor equal to $U$. Such sets are called IP (for Indecomposable Past set), and those that are not of the form $U=I^-(x)$ for some $x\in M$ are called TIP's (for Terminal IP's);  these TIP's are $U=I^-(\gm)$ for some  future-endless timelike curve $\gm$. It would  be more natural if strong cosmic censorship merely excluded visible TIP's coming from incomplete curves, but Penrose (1979) gives various arguments for including complete curves $\gm$, too, and in any case his Theorem 
 \ref{P79theorem} below holds only if all TIP's are included.  }
 
\noindent   Now, if $\gm$ does have a future endpoint $y\neq x$, the crucial condition 
$I^-(\gm)\subset I^-(x)$
  occurring in Definition \ref{CSdef1} below--albeit in the endless case--is evidently equivalent to $I^-(y)\subset I^-(x)$, i.e.\ $y\ll x$, which  states that there exists an fd \emph{timelike} curve or signal from $y$ to $x$. If $\gm$ is endless, on the other hand, there is no such point $y$, but we may still interpret \er{Isubset} as saying that  timelike signals emanating from the ideal point $y$ defined by $\gm$ (such as a singularity), or from arbitrarily nearby points, can reach $x$.  
  This exegesis also applies to the condition  $I^-(\gm)\subset J^-(x)$, in which case some \emph{causal} curve from $y$  reaches $x$. 
  
  The following definition then captures the two notions of cosmic censorship in Penrose (1979).\footnote{See also  Kr\'{o}lak (1986) for a  different  axiomatization of weak cosmic censorship \`{a} la Penrose, as well as useful analysis.}
   We recall that these definitions and the ensuing theorem presuppose that $(M,g)$ is strongly causal.\footnote{Strong causality is used through its implication that $I^{\pm}(x)=I^{\pm}(y)$ iff $x=y$, without which Definition \ref{CSdef1} would make little sense, as well as through its implication of non-total imprisonment, without which the invocation of Theorem 2.53 in Minguzzi (2019) in the proof of Theorem \ref{P79theorem} below would fail. It is also part of one of the traditional definitions of global hyperbolicity (namely strong causality plus compactness of causal diamonds $J^+(y)\cap J^-(x)$), but since the proof of  Theorem \ref{P79theorem} is based on contradicting compactness of causal diamonds, in that role it is hardly necessary anymore, since Hounnonkpe \& Minguzzi (2019) proved that  a non-compact space-time with $\dim(M)\geq 3$ is globally hyperbolic iff all  double cones are compact.}
   \begin{definition}\label{CSdef1}
In both cases below, let $\gm$ denote a future-directed future-endless causal curve.\footnote{There is a similar definition in terms of  \emph{past-directed} endless causal curves, in which $I^-(\cdot)$ is replaced by  $I^+(\cdot)$ throughout. As far as strong cosmic censorship is concerned this definition turns out to be equivalent to the given one, cf.\ Theorem \ref{P79theorem} below, whilst for  weak cosmic censorship the above definition is the appropriate one. Note that strong cosmic censorship does not imply weak cosmic censorship since \er{Isubset0} has $J^-(x)$ with $x$ possibly in $\CI^+\subset\p\til{M}$, whilst \er{Isubset} has $I^-(x)$ with $x\in M$.
 } 
  \begin{itemize}
  \item  A space-time $(M,g)$ that is asymptotically flat at null infinity (\S\ref{ni}) contains a \hi{naked singularity} if  there is a curve $\gm$ as above, and a point $x\in J^-(\CI^+)$ in its causal future in $\til{M}$,  in the sense that 
  \beq
  I^-(\gm)\subset J^-(x).  \label{Isubset0}
  \eeq
   Penrose's \hi{weak cosmic censorship conjecture} states that  space-times that are asymptotically flat at null infinity and arise from  ``ge\-ne\-ric''  regular initial conditions contain no naked singularities.
\item A space-time $(M,g)$  contains a \hi{locally naked singularity} if there is a curve $\gm$ as above, and a point $x\in M$  in its chronological future, in the sense that
 \beq
I^-(\gm)\subset I^-(x). \label{Isubset}
\eeq
 Penrose's \hi{strong cosmic censorship conjecture} states that ``ge\-ne\-ric'' [in his own words: ``physcially reasonable''] space-times do not contain locally naked singularities.
\end{itemize}
   \end{definition}
It should be defined precisely what ``generic''  means, lest these conjectures turn into a definition of genericity! Penrose did not do this, and we will return to this point in \S\ref{PDECCC}. It is important to realize that in this definition Penrose does not require $\gm$ to be \emph{incomplete}, but merely \emph{endless}. Indeed the notion of (in)completeness is hard to define for non-geodesic curves since it depends on the parametrization;  if, as we do, continuous causal curves are parametrized by arc length as measured by an auxiliary complete Riemannian metric (see footnote \ref{hfn}), then the distinction between endlessness and incompleteness cannot even be made, because any endless curve has infinite arc length.\footnote{Recall that affinely parametrized geodesics are incomplete iff they are endless and have finite parameter length.} Beyond moving from causal geodesics to general  causal curves, this further generalization allows even more singularities, and has the effect of making the notion(s) of cosmic censorship more stringent--in excluding a larger class of naked or locally naked singularities--than Penrose's (1965)  singularity theorem would suggest.\footnote{The conditions \er{Isubset0}
and \er{Isubset} do make  sense for complete future endless causal curves: for example,  in anti-de Sitter space one has   endless causal curves $\gm$ and points $x$ such that $I^-(\gm)\subset J^-(x)$, but this space is not regarded as singular (it is a  space-time of constant negative curvature).  
Penrose (1979), p.\ 623 notes that this is impossible in space-times that are asymptotically flat at null infinity, and indeed
 anti-de Sitter space has a negative cosmological constant with timelike future null infinity. 
 }

The following theorem, of which part 2 is due to Penrose (1979) with a slightly different proof, and part 1 is an almost trivial addition, is the main characterization of cosmic censorship in his sense.  
 \begin{theorem}\label{P79theorem} 
\begin{enumerate}
\item  If $(M,g)$ is asymptotically flat at null infinity, then it has no  naked singularities iff the exterior region  $J^-(\CI^+)$ in $\til{M}$ (which by definition includes $\CI^+\subset\p\til{M}$)
is globally hyperbolic.\footnote{Tipler, Clarke \& Ellis (1980), p.\ 176,  made this the
\emph{definition} of weak cosmic censorship. It may be closer to Penrose's (1979) formulation to require 
$x\in J^-(\CI^+)\cap J^+(\Sg)$ in the first part of Definition \ref{CSdef1}, where $\Sg$ is some partial Cauchy surface in $M$,
in which case Theorem \ref{P79theorem} yields global hyperbolicity of $J^-(\CI^+)\cap J^+(\Sg)$. This is similar to the condition  $\CI^+\subset\ovl{D^+(\Sg)}$ making
 $(M,g)$  \emph{future asymptotically predictable} from $\Sg$ (Hawking \& Ellis, 1973, p.\ 312), but is equivalent to it only under further regularity assumptions (Kr\'{o}lak, 1986, Lemma 2.10). See also Wald (1984), \S 12.1 and Chru\'{s}ciel (2020), \S 3.5.1.
  }
\item In general, a space-time $(M,g)$ has no locally naked singularities iff it is globally hyperbolic.
\end{enumerate}
\end{theorem}
It should be clear intuitively that at least part 2 of the theorem is true (in the contrapositive): if a space-time contains a locally naked singularity, represented by $\gm$ as in Definition \ref{CSdef1}, then $\gm$ will not reach any partial Cauchy-surface $\Sg$ lying in the future of $x$, since it crashes at the singularity lying in the past of $x$. Conversely, if no Cauchy surface exists then one can construct such a curve $\gm$. See also \S\ref{examples}.  \smallskip

\noindent 
\emph{Proof.}  We  prove the inference from a locally naked singularity to non-global hyperbolicity by contradiction.  Suppose that \er{Isubset} holds for some $\gm$ and $x$ and that $(M,g)$ is globally hyperbolic.
 Take $y\in \gm$ and then a future-directed sequence $(y_n)$ of points on $\gm$, with $y_0=y$. 
 Because of \er{Isubset} this sequence lies in  $J^+(y)\cap J^-(x)$, which is compact by assumption. Hence 
 $(y_n)$ has a limit point $z$ in  $J^+(y)\cap J^-(x)$. Now define curves $(c_n)$ as the segments of $\gm$ from $y$ to $y_n$.
 By the curve limit lemma,\footnote{One needs Theorem 2.53 in Minguzzi (2019), of which part (i) applies:
 Let $(c_n: [0,b_n]\raw M)$ be a sequence of fd continuous causal curves parametrized by $h$-arc length in a non-imprisoning space-time such that $c_n(0)\raw x$ and $c_n(b_n)\raw y\neq x$. Then there exists an fd continuous causal curve $c:[0,b]\raw M$,
where $b<\infty$ as well as a subsequence of $(c_n)$ that converges uniformly to $c$ (including
$b_n\raw b$ at the endpoint).
  Penrose (1979) gives a more complicated argument, perhaps since the version of the  curve limit lemma just cited was not available at the time, or because he wanted to use his TIP's (which we avoid).} these curves have a uniform limit, whose arc length (as measured by an auxiliary complete Riemannian metric, see footnote \ref{hfn})
  is on the one hand infinite (since $\gm$ is endless and hence has infinite arc length, which is approached as the $y_n$ move up along $\gm$), but on the other hand is finite, since it must end at $z$ (and fd continuous causal curves have finite arc length iff they have an endpoint). 
   Hence $(M,g)$ cannot be globally hyperbolic.\footnote{This also works for part 1, where  $x\in J^-(\CI^+)$), eq.\  \er{Isubset0} also implies $y_n\in  J^+(y)\cap J^-(x)$.
    Conversely, 
if there are $x,y$ for which $J^-(x)\cap J^+(y)$ is not compact, one can easily construct  a future-directed future-endless causal curve $\gm$ such that \er{Isubset} holds for the given $x$ (Penrose, 1979, p.\ 624). Since  \er{Isubset} trivially gives $I^-(\gm)\subset J^-(x)$, also this implication works for  part 1.}
 \enp\smallskip

Especially its definition through the existence of a Cauchy surface relates global hyperbolicity to \emph{determinism}, the idea being that any event in a globally hyperbolic space-time is determined by certain initial data on a Cauchy surface in it, at least as long as the (classical) universe is governed by hyperbolic partial differential equations. This is clearly true for the gravitational field itself (as long as it satisfies the Einstein equations), and also has considerable backing for other fields.\footnote{See e.g.\  Choquet-Bruhat (2009) and B\"{a}r, Ginoux, and Pf\"{a}ffle (2007), respectively, as well as Earman (1995, 2007).}
This does not imply that \emph{non}-globally hyperbolic space-times are necessarily \emph{in}deterministic:  the point is rather that signals from a (locally) naked singularity can reach an event without ultimately coming from a 
Cauchy surface, so that the event is influenced by data other than those at an initial-value surface. Thus the event in question may still be fully determined--but it is not determined by the initial data that were supposed to do so.\footnote{
The closest analogue to this generally relativistic situation occurs in \emph{non-relativistic} mechanics, where bodies may disappear to infinity in finite time (Xia, 1992; Saari \& Xia, 1995), and hence, by the same (time-reversed) token, may \emph{appear} from nowhere in finite time and hence influence affairs in a way unforeseeable from any Cauchy surface.
See Earman (2007), \S 3.6.}

Conversely,  the flagrant indeterminism concerning the unknown fate of someone falling into a black hole singularity is compatible with global hyperbolicity (as in e.g.\ the Schwarzschild solution). Furthermore, suppose some globally hyperbolic space-time $(M,g)$ solving the Einstein equations is metrically extendible (see \S\ref{sst}), such that the extension $(M',g')$  is either not globally hyperbolic at all, or is  globally hyperbolic but not with respect to any Cauchy surface $\Sg$ in $M$. Then again, although all things in $(M',g')$ may be determined, they are not determined by the initial data on $\Sg$ one expected to do so.\footnote{Doboszewski (2017, 2019, 2020) analyzes
 the connection between global hyperbolicity, extendibility, and determinism.} 
\subsection{Cosmic censorship in the initial value (\pde) formulation}\label{PDECCC}
Definition \ref{CSdef1} of the cosmic censorship conjectures is inappropriate from the point of view of the initial-value problem. 
Recall from \S\ref{HI}  that in this approach all valid questions are  about the maximal globally hyperbolic development $(M,g,\iota)$ of  initial data $(\Sg,h,K)$. Since the \mghd\ is always globally hyperbolic, the strong version is trivial by Theorem \ref{P79theorem}. For weak cosmic censorship there is a subtle issue about which asymptotically flat  initial data lead to 
\mghd\ that are  asymptotically flat at null infinity (see footnote \ref{fn14}), but even granting this, the real problem is that 
 even in clear counterexamples to the Penrosian conjecture (such as $m<0$ Schwarzschild, see \S\ref{examples}), the 
 space $J^-(\CI^+)$ computed for the \mghd\ is  globally hyperbolic. Thus the Penrosian version, applied to the \mghd, would 
 hold despite naked singularities!

There is no  crystal-clear logical path from Penrose's formulation of the  cosmic censorship conjectures to the current versions used in the \pde\ literature, but there is some continuity of ideas. First, as to the weak version, in order  to 
 strengthen Definition \ref{defAF} (or rather some slight variation thereof) 
  Geroch \& Horowitz (1978) proposed  that  \emph{future null infinity $\CI^+$ be null geodesically complete}.\footnote{
  \label{GHF} This condition is nontrivial to state; for example, in the metric $\til{\eta}$ used below even $\CI^+$ for the standard 
  conformal completion $(\til{\Mi},\til{\eta})$ of Minkowski space-time $(\Mi,\eta)$ is incomplete. Completeness of curves depends on their parametrization. Geodesics are affinely parametrized by definition (and an affine reparametrization does not affect their (in)completeness), but a change in $\Om$ changes the unphysical metric $\til{g}$ (for given physical metric $g$). Hence  the notion of a geodesic and its  (in)completeness  depends on the  choice of $\Om$. 
  As recognized  by  Geroch \& Horowitz (1978) themselves, the correct approach is to use the freedom of rescaling $\Om$ to ensure that $\til{\n}_{\mu}\til{\n}_{\nu}\Om=0$ on $\CI^+$, and require null geodesic completeness of the null hypersurface  $\CI^+$ in this ``gauge'', in which the flow of $\til{\n}\Om$ is geodesic. See also Wald (1984), \S 11.1 or Stewart (1991), \S 3.6. }
   This was motivated by the following example. In  light-cone coordinates, the standard conformal completion $(\til{\Mi},\til{\eta})$ of Minkowski space-time $(\Mi,\eta)$,
  described for example in Penrose (1968), pp.\ 175--177, is given by 
  \begin{align}
\hat{\Mi}&:=\{(p,q,\theta,\varphi)\mid (p,q)\in (-\half\pi,\half\pi)^2, p\geq q, (\theta,\varphi)\in S^2\}\cup\CI^+\cup\CI^-;
\label{defhatMi} \\
\CI^+&=\{(p,q,\theta,\varphi)\mid p=\half\pi, q\in (-\half\pi,\half\pi), (\theta,\varphi)\in S^2\};\label{CIp1}\\
\CI^-&=\{(p,q,\theta,\varphi)\mid p\in (-\half\pi,\half\pi),q=-\half\pi, (\theta,\varphi)\in S^2\};\label{CIm1}\\
\hat{\eta}&=-dp\, dq+\quar\sin^2(p-q)(d\theta^2+\sin^2\theta d\phv^2); \label{hatetaP}\\
 \Om&=\cos p\cos q.
\end{align}
 Now on the one had,  truncating $\CI^+$ to for example $\{(p,q,\theta,\phv)\mid p=\half\pi, q\in (-\half\pi,0)\}$ instead of \er{CIp1} would still define a conformal completion of $(\Mi,\eta)$, with respect to which the future light-cone $J^+(0)$ is a fake black hole $\mathcal{B}$ in $\Mi$. On the other hand, removing $\mathcal{B}=J^+(0)$, the ensuing space-time $(\Mi\backslash J^+(0),\eta)$  has a conformal completion (such as the one just described), which by design is free of black holes. In both undesirable cases  future null infinity is incomplete (in the  sense of footnote \ref{GHF}).\footnote{This clause seems to be an improvement over an inextendibility condition proposed by Geroch (1977), which did not exclude cases like $(\Mi\backslash J^+(0),\eta)$. However, inextendibility plus some regularity condition enabled Geroch (1977) to prove uniqueness of conformal completions, a result that seems to have no analogue for Definition \ref{defAF}, even in strengthened form.
}

Completeness of future null infinity in the above sense, then, was  taken to be the 
 \pde\ reformulation of weak cosmic censorship  (Christodoulou, 1999a), although it turns the  Penrosian version on its head! For whereas his version states that \emph{outgoing} signals from a black hole singularity are blocked by an event horizon $\mathcal{H}^+_E$, the new version is about \emph{incoming} (null) signals: the further these are away from $\mathcal{H}^+_E$, the longer it takes them to enter $\mathcal{H}^+_E$, and in the limit at null infinity this takes infinitely long, making $\CI^+$ complete. 
Yet the \pde\ version appears to strengthens Penrose's: heuristically, and
 contrapositively,  lack of global hyperbolicity of $J^-(\CI^+)$ gives a partial Cauchy surface $\Sigma$ a Cauchy horizon which cuts off
 $\CI^+$, making it incomplete (this will perhaps be clearer from the examples in \S\ref{examples}).\footnote{Christodoulou (1999a) actually reformulates the above definition of weak cosmic censorship in such a way that the idealization $\CI^+$ no longer occurs.
 Let $(\Sg,h,K)$ be asymptotically flat initial data for the Einstein equations (satisfying the constraints),  with \mghd\ $(M,g,i)$. He then defines $(M,g)$  to have ``complete future null infinity'' iff for any $s>0$ there exists a region $B_0\subset B\subset\Sg$ such that $\p D^+(B)$, which is ruled by null geodesics, has the property that each null geodesic starting 
 in $\p J^+(B_0)\cap \p D^+(B)$ can be future extended beyond parameter value  $s$. 
 Here $D^+(B)$ is the future domain of dependence of $B$,  and each null geodesic in question is supposed to have tangent vector $L=T-N$, where $T$ is the fd unit normal to $\Sg$ in $M$ and $N$ is the outward unit normal to $\p B$ in $\Sg$.
 See also Christodoulou \& Klainerman (1993) for background on these constructions. } Thus we obtain:
\begin{definition}\label{CSfed1}
\begin{itemize}
\item The \hi{weak cosmic censorship conjecture} states that if ``generic'' complete initial data have a \mghd\ that is asymptotically flat at null infinity, then  future null infinity is  complete.
\item The \hi{strong cosmic censorship conjecture} states that the \mghd\ of ``generic'' complete initial data  is metrically inextendible (as a space-time in a regularity class to be specified in detail).\end{itemize}
\end{definition}
For convenience, we have  added the strong version of cosmic censorship used in the \pde\ approach, whose path 
from the Penrosian formulation we now try to trace.\footnote{I am greatly indebted to Juliusz Doboszewski for drawing my attention to the early papers by Moncrief \emph{et al}.}
First, in a paper on \emph{weak} cosmic censorship, Moncrief \& Eardley (1981), p.\ 889, propose an `(informally stated) global existence conjecture':
\begin{quote}
\begin{small}
Every asymptotically flat initial data set with $\mathrm{tr} K=0$ may be evolved to arbitrarily large times (\ldots)
\end{small}
\end{quote}
adding that its proof would `in essence prove the [weak] cosmic censorship conjecture for asymptotically flat space-times'. 
For initial data given on a compact Cauchy surface they  propose something similar, and in doing so they opened the door to regarding cosmic censorship as a global existence problem for the (vacuum) Einstein equations, as indeed the title of their paper already expresses. In this spirit, Moncrief (1981), p.\ 88, paraphrases  Penrose's strong version as expressed by Theorem \ref{P79theorem} as 
\begin{quote}
\begin{small}
i.e., that the maximal Cauchy development of a generic initial data set is inextendible.
\end{small}
\end{quote}
This is  made more precise by Chru\'{s}ciel,  Isenberg,  \& Moncrief (1990), who open their abstract as follows:
\begin{quote}
\begin{small}
The strong cosmic censorship conjecture states that `most' spacetimes developed as solutions of Einstein’s equations from prescribed initial data cannot be extended outside of their maximal domains of dependence. 
\end{small}
\end{quote}
They later (\S 3) specify the word `most' in terms of open and dense subsets in the space of initial data.\footnote{Detailed mathematical criteria for genericity (which are suggested by \pde\ theory and whose physical relevance is  doubted outside the \pde\ community) may be also be found for example in Dafermos (2003) and 
Luk \& Oh (2019a), \S 3.}

Chru\'{s}ciel (1992) introduced the notion of a 
 \emph{development} of initial data $(\Sg,h,K)$  as a triple $(M,g,i)$, where $(M,g)$ is a $4d$ space-time solving the vacuum Einstein equations, and $i:\Sigma\raw M$ is an embedding such that $i^*g=h$ and $i(\Sg)$ has extrinsic curvature  $K$;
 the difference from a Cauchy development (see  footnote \ref{CGBfn}) is that 
 $i(\Sg)$ is no longer required to be Cauchy surface in $M$, so that $(M,g)$ is not necessarily globally hyperbolic.
 He calls such a development \emph{maximal} if there is no  extension $(M',g')$ that also satisfies the vacuum
Einstein equations, and proves existence of maximal developments (but not uniqueness up to isometry, as in  the globally hyperbolic case, cf.\ footnote \ref{CGBfn}). Applying Penrose's strong cosmic censorship to such a maximal development, he  asks it to be globally hyperbolic. If this is the case, then--up to isometry as usual--$(M,g)$  must coincide with the  \mghd\ of  given initial data.\footnote{Continuing footnote \ref{CGBfn}, the set of isometry classes $[M,g,i]$ of Cauchy developments $(M,g,i)$ of given initial data $(\Sigma, h,K)$ is partially ordered by $[M_1,g_1,i_1]\leq [M_2,g_2,i_2]$ provided there are representatives $(M'_1,g'_1,i'_1)$ and $(M'_2,g'_2,i'_2)$  and an embedding $\psi:M'_1\raw M'_2$ for which $\psi^*g'_2=g'_1$ and $\psi\circ\iota'_1=\iota'_2$. The \mghd\ $[M_t,g_t,i_t]$ is the top element of this poset (Sbierski, 2016) and so
if some maximal development  $(M_m,g_m,i_m)$ \`{a} la Chru\'{s}ciel is globally hyperbolic then 
$[M_m,g_m,i_m]\leq [M_t,g_t,i_t]$. On the other hand, since $(M_t,g_t,i_t)$ is a solution and $(M_m,g_m,i_m)$ is
maximal also the converse holds, so $(M_m,g_m,i_m)\cong  (M_t,g_t,i_t)$.}
 Consequently, this specific application of 
strong cosmic censorship \`{a} la Penrose is equivalent to
 asking  the \mghd\ of  given initial data to be  inextendible \emph{as a solution to the vacuum (or any kind of) Einstein equations}.\footnote{See  Doboszewski (2017, 2019) and Manchak (2011, 2017) for conceptual studies of the (in)extendibility of space-times. }
 
Adding suitable regularity conditions on the  extensions,\footnote{As pointed out to me by Juliusz Doboszewski,
Chru\'{s}ciel, Isenberg, \& Moncrief (1990) as well as Chru\'{s}ciel \& Isenberg (1993) only consider smooth extensions,
so that looking at lower regularity seems a refinement postdating this early phase.} 
 this would be a meaningful and natural \pde\ version of strong cosmic censorship  but the version used in the \pde\ literature is stronger: one requires metric inextendibility of the \mghd\ full stop, \emph{whether or not this extension satisfies the vacuum Einstein equations}. And although it would make sense in general, 
 in practice the ensuing conjecture is posed for either non-compact $\Sg$ with asymptotically flat initial data or compact $\Sg$; better safe than sorry!

 The need for 
a restriction on the scope of the conjectures was clearly realized and stated--albeit purely qualitatively--already by Penrose himself (see footnote \ref{PD}). 
Indeed,  without such a restriction some of the best-known exact black hole solutions (cf.\ \S\ref{examples})  provide  counterexamples to one or both of the conjectures, as was of course well known to Penrose and his circle (for the Penrosian version, that is).
To get around this,
in one of his most prophetic insights, 
Penrose (1968, p.\ 222) suggested that  because of a blueshift instability of the Cauchy horizon under perturbations,  it  turns into a curvature singularity:
\begin{quote}
\begin{small}
Our contention in this note is that if the initial data is generically perturbed then the Cauchy horizon does not survive as a non-singular hypersurface. It is strongly implied that instead, genuine space-time singularities will appear along the region which would otherwise have been the Cauchy horizon. (Simpson \& Penrose, 1973, p.\ 184)
\end{small}
\end{quote}
Since then, this  instability has  been confirmed in a large number of studies, starting with Hiscock (1981) in the physics literature and  Dafermos (2003) in the  mathematical one; recent papers include Chesler,   Narayan \&  Curiel (2020)  and Van de Moortel (2020), respectively.
 The conclusion seems to be that Cauchy horizons turn into so-called \emph{weak null singularities},\footnote{These are null boundaries with $C^0$  metric  but  Christoffel symbols not locally in $L^2$ (Luk \& Sbierski, 2016; Luk, 2017).}
  behind which--at least for one-ended asymptotically flat initial data--there is a strong curvature singularity at $r=0$. See also Luk \& Oh (2019ab) for the two-ended case. Unappealingly,
  the sense in which strong cosmic censorship (in the \pde\ formulation) then fails or holds depends critically on the regularity assumptions of the extension.\footnote{The results below concern cosmological constant  $\lm=0$ and subextremal black holes (i.e.\ $e^2<m^2$ for R--N and $a^2<m^2$ for Kerr). 
  See Dias, Reall, \& Santos (2018) for   $\lm>0$:  strong cosmic censorship seems true in pure gravity and false for the Einstein--Maxwell system, but again this  depends critically on the regularity of the extension. For extremal Reissner--Nordstr\"{o}m ($e^2=m^2$) at $\lm=0$ see Gajic \& Luk (2017), suggesting failure of strong cosmic censorship, as is trivially the case for $e^2>m^2$.} 
  
  For example, for two-ended asymptotically flat data for the spherically symmetric Einstein--Max\-well-scalar field system (to which the  conjecture, so far discussed for the vacuum case, can be extended in the obvious way), the strong cosmic censorship conjecture \emph{fails} in $C^0$ (Dafermos \& Luk, 2017),\footnote{Their general result assumes the (widely expected) stability of the Kerr metric under perturbations of the initial data.}
   but it \emph{holds} in $C^0$ with the additional requirement that the associated Christoffel symbols are locally $L^2$ (Luk \& Oh, 2019ab). This is not just a technicality, since
having the metric in $C^0$  and its Christoffel symbols  locally  $L^2$ is a borderline
   regularity condition for metric extensions in strong cosmic censorship: it is the least regular case in which the metric 
    can still be  defined as a weak solution to Einstein's equations (Christodoulou, 2009, p.\ 9; Luk, 2017, footnote 1).
    Indeed, a weak solution of the vacuum Einstein equations is a metric $g$ for which  for all compactly supported $X,Y\in\XM$,
\beq
\int_M d^4x\, \sqrt{-\det(g(x))} R_{\mu\nu}(x)X^{\mu}(x)Y^{\nu}(x)=0.
\eeq
 Partial integration shows that this is well defined iff the $\Gm_{\mu\nu}^{\rh}$ are  locally $L^2$. This simple observation should not be confused with the very deep result that having 
 the \emph{Ricci tensor} in $L^2$ is sufficient for the (vacuum)  Einstein equations to be weakly \emph{solvable} at least locally
(Klainerman,   Rodnianski, \& Szeftel, 2015). Ironically, in  Definition \ref{CSfed1} of strong cosmic censorship 
 the extension is not required to satisfy the Einstein (or indeed any other) equations! 
  See  also Ringstr\"{o}m (2009) for a review of the 
`cosmological'  case  where the Cauchy surface is 
 compact, in which the strong (\pde) conjecture seems to hold.

 The status of weak cosmic censorship is even less clear. Christodoulou (1999b)  proves the conjecture for the spherically symmetric gravitational collapse of a scalar field, but on the basis of genericity conditions whose  relevance has been questioned in the physics literature (Gundlach \& Martin-Garcia, 2007, \S 3.4). More generally, the status of weak cosmic censorship seems mixed also in earlier heuristic formulations in terms of  an event horizon; see e.g.\ Joshi (1993, 2007),  Kr\'{o}lak (1999), and Ong (2020).
 \section{Examples}\label{examples}
The relationship between the Penrosian and the \pde\ versions of the cosmic censorship conjectures is best understood from  three key black hole examples and their Penrose diagrams:\footnote{Even more so than the previous sections this one is purely pedagogical and drawn largely from  Hawking \& Ellis (1973), pages 158 and 160, as well as from Dafermos \& Rodnianski (2008) and Dafermos (2013, 2014ab, 2017, 2019) for the \pde\ side.}
\begin{itemize}
\item  Maximally extended Schwarz\-schild  (i.e.\ Kruskal) with $m>0$ (and two-sided initial data);
\item  Schwarz\-schild  with  $m<0$, which in so far as singularities and horizons are concerned also looks like supercharged Reissner--Nordstr\"{o}m  ($e^2>m^2>0$), or ultrafast rotating Kerr ($a^2>m^2>0$);
\item  Reissner--Nordstr\"{o}m with $0<e^2<m^2$, which  qualitatively also represents Kerr with $0<a^2<m^2$.
\end{itemize}
In the first case the  solution coincides with the \mghd\ of the corresponding (two-ended) initial data, so  the difference between the Penrosian and the \pde\ approach evaporates. Here is the Penrose diagram:
\begin{center}
   \begin{tikzpicture}
     \draw[fill=gray!10]    (-6,0) -- ++(3,3) -- ++(6,0) -- ++(3,-3)-- ++(-3,-3)-- ++(-6,0) -- ++(-3,3) ;   
         \draw[blue,thick] (-6,0) -- (6,0);
            \draw[green,thick] (-3,-3) -- (3,3);
                                                \draw[green,thick] (-3,3) -- (3,-3);
                                                \draw[decorate,decoration=zigzag]  (-3,3) -- (3,3)
    node[midway, above, inner sep=2mm] {$r=0$};
      \draw[decorate,decoration=zigzag]  (-3,-3) -- (3,-3)
    node[midway, below, inner sep=2mm] {$r=0$};
    \draw (3,3) -- (6,0);
\draw (3,-3) -- (6,0);
\draw (-3,-3) -- (-6,0);
\draw (-3,3) -- (-6,0);
          \node[label=above:$\mathcal{H}_E^+$]  at (0,0.25) {};
                                     \node[label=below:$\mathcal{H}_E^-$]  at (0,-0.25) {};
                                         \node[label=above:II]  at (0,1.5) {};
                                     \node[label=below:IV]  at (0,-1.5) {};
                                          \node[label=above:$\Sigma$]  at (-3,0) {};
                                                                  \node[label=above:$\Sigma$]  at (3,0) {};
                                                                    \node[label=below:III]  at (-3,-1) {};
                                                                      \node[label=above:III]  at (-3,1) {};
                                                                      \node[label=below:I]  at (3,-1) {};
                                                                       \node[label=above:I]  at (3,1) {};
                                     \node[label=right:$\mathcal{I}^+$]  at (4.5,1.5) {};
                                 \node[label=left:$\mathcal{I}^+$]  at (-4.5,1.5) {};    
                      \node[label=right:$\mathcal{I}^-$]  at (4.5,-1.5) {};
                                 \node[label=left:$\mathcal{I}^-$]  at (-4.5,-1.5) {};      
                                       \node[label=right:$i^0$]  at (6,0) {}; 
                                           \node[label=left:$i^0$]  at (-6,0) {}; 
                        \node[label=above:$i^+$]  at (3,3) {};
                          \node[label=above:$i^+$]  at (-3,3) {};
         \node[label=below:$i^-$]  at (3,-3) {};
                          \node[label=below:$i^-$]  at (-3,-3) {};                                               
    \end{tikzpicture}
\end{center}
\emph{Penrose diagram of the maximally extended Schwarzschild solution with $m>0$. This solution coincides with the maximal Cauchy development (marked in light grey) of a generic two-sided Cauchy surface $\Sg$ with suitable initial data, drawn as a horizontal blue line.  Thus the Cauchy horizon $\mathcal{H}^{\pm}_C$ is empty. The upper two green lines form the future event horizon $\mathcal{H}^+_E$ of the black hole area, which is the upside-down upper triangle (labeled region II), whereas the lower two green lines form the past event horizon $\mathcal{H}^-_E$ of the white hole area, i.e.\ the  lower triangle (region IV). The right-hand diamond is region I, the left-hand diamond is region III.
Fd causal curves cannot \emph{leave} region II and they cannot 
\emph{enter} region IV.
}\smallskip

 Both cosmic censorship conjectures hold in both versions (i.e.\ Penrose and \pde):
\begin{itemize}
\item  \emph{Weak  cosmic censorship.} Penrose:  $\Sg$ is a Cauchy surface for $J^-(\CI^+)$, making it globally hyperbolic.\footnote{Alternatively: any \emph{incomplete} future inextendible timelike curve $\gm$  must crash in the upper $r=0$ singularity. Hence $I^-(\gm)$ lies partly in region II, which is disjoint from $J^-(\CI^+)$, so that $I^-(\gm)\nsubseteq J^-(x)$  for all $x\in J^-(\CI^+)$.} \pde: each component of $\CI^+$ ends  at timelike infinity and hence all its null geodesics are future complete (as confirmed by explicit parametrization and computation). 
\item \emph{Strong  cosmic censorship.}   Penrose: Kruskal space-time is globally hyperbolic (since the causal structure
of the diagram  is such that the line $\Sg$ represents a Cauchy surface).
 \pde: For smooth extensions
 Remark 5.45 on page 155 of O'Neill (1983) or Proposition 4.4.3 in  Chru\'{s}ciel (2020) plus a detailed study of the geodesics shows that  Kruskal space-time is metrically
 inextendible.\footnote{ If for any maximally extended timelike geodesic $\gm:[0,b)\raw M$ in $M$ there is a curvature invariant (such as $R$ or $R^{\rh\sg\mu\nu}R_{\rh\sg\mu\nu}$, etc.)
  that blows up as $\gm(t)\raw b$, then $(M,g)$ is  inextendible. 
 See O'Neill (1983), Chapter 13, for a study of Kruskal geodesics,  proving the antecedent.
Sbierski (2018ab) proves that Kruskal space-time is inextendible even in $C^0$.}
\end{itemize}

 \noindent However, for $m<0$ Kruskal, Reissner--Nordstr\"{o}m, and Kerr, differences arise between the Penrosian and the \pde\ perspectives, since in these cases the  maximal (analytic) solutions, deemed unphysical by the \pde\ aficionados, differ from the \mghd\ of the pertinent initial data. In particular, although (curvature) singularities are not part of space-time in any case,
they can at least be drawn as boundaries in the maximal solutions, where they lie behind a Cauchy horizon. But precisely for that reason  singularities are outside any kind of scope of the corresponding \mghd.  Here are the Penrose diagrams:
\begin{center}
\begin{minipage}{0.4\textwidth}
  \centering
   \begin{tikzpicture}
\draw (0,-6) -- (6,0);
\draw (6,0) -- (0,6);
\draw[decorate,decoration=zigzag]  (0,-6) -- (0,6)
    node[midway, left, inner sep=2mm] {$r=0$};
       \node[label=right:$\mathcal{I}^+$]  at (3,3) {};
            \node[label=right:$\mathcal{I}^-$]  at (3,-3) {};
               \node[label=left:$\mathcal{H}_C^+$]  at (2.0,2.0) {};
               \node[label=left:$\mathcal{H}_C^-$]  at (2.0,-2.0) {};
                     \node[label=right:$i^0$]  at (5.75,0) {}; 
                        \node[label=above:$i^+$]  at (0,5.75) {};
                         \node[label=below:$i^-$]  at (0,-5.75) {};  
                  \draw[fill=gray!10]    (0,0) -- ++(3,3) -- ++(3,-3) -- ++(-3,-3)-- ++(0,-0);                            
            \draw[red,thick] (0,0) -- (3,3);
             \draw[red,thick] (0,0) -- (3,-3);
               \draw[blue,thick] (0,0) -- (6,0);
                    \node[label=above:$\Sigma$]  at (3,0) {};
    \end{tikzpicture}
\end{minipage}%
\begin{minipage}{0.1\textwidth}
\mbox{} \end{minipage}%
\begin{minipage}{0.4\textwidth}
  \centering
   \begin{tikzpicture}
    \draw[fill=gray!10]    (-4,0) -- ++(4,4) -- ++(4,-4) -- ++(-4,-4)-- ++(-4,4);   
   \draw[decorate,decoration=zigzag]  (2,2) -- (2,6)
    node[midway, right, inner sep=2mm] {$r=0$};
    \draw[decorate,decoration=zigzag]  (-2,2) -- (-2,6)
    node[midway, left, inner sep=2mm] {$r=0$};
       \draw[decorate,decoration=zigzag]  (2,-2) -- (2,-6)
    node[midway, right, inner sep=2mm] {$r=0$};
    \draw[decorate,decoration=zigzag]  (-2,-2) -- (-2,-6)
    node[midway, left, inner sep=2mm] {$r=0$};
               \draw[red,thick] (2,2) -- (-2,6);
                    \draw[red,thick] (-2,2) -- (2,6);
                               \node[label=above:$\mathcal{H}_E^+$]  at (0,0.25) {};
                                     \node[label=below:$\mathcal{H}_E^-$]  at (0,-0.25) {};
                                    \draw[blue,thick] (-4,0) -- (4,0);
                                          \draw[green,thick] (-2,2) -- (0,0);
                                                \draw[green,thick] (2,2) -- (0,0);
                                                         \node[label=above:$\Sigma$]  at (-2,0) {};
                                                                  \node[label=above:$\Sigma$]  at (2,0) {};
                                    \draw (-2,2) -- (-4,0);{};  
                                       \draw (2,2) -- (4,0);{};    
            \node[label=below:$\mathcal{H}_C^+$]  at (0,3.75) {};
                      \node[label=right:$\mathcal{I}^+$]  at (2.75,1.25) {};
                                 \node[label=left:$\mathcal{I}^+$]  at (-2.75,1.25) {};    
               \node[label=above:$\mathcal{H}_C^-$] at (0,-3.75) {};
                      \node[label=right:$\mathcal{I}^-$]  at (2.75,-1.25) {};
                                 \node[label=left:$\mathcal{I}^-$]  at (-2.75,-1.25) {};    
                \draw[red,thick] (-2,-2) -- (2,-6);
                    \draw[red,thick] (2,-2) -- (-2,-6);
                     \draw[green,thick] (-2,-2) -- (0,0);
 \draw[green,thick] (2,-2) -- (0,0);         
            \draw (-2,-2) -- (-4,0);{};  
                                       \draw (2,-2) -- (4,0);{};   
                     \node[label=right:$i^0$]  at (6,0) {}; 
                        \node[label=above:$i^+$]  at (2.4,1.8) {};
                          \node[label=above:$i^+$]  at (-2.4,1.8) {};
         \node[label=below:$i^-$]  at (2.4,-1.8) {};
                          \node[label=below:$i^-$]  at (-2.4,-1.8) {};   
                                     \node[label=right:$i^0$]  at (3.8,0) {};
 \node[label=left:$i^0$]  at (-3.8,0) {};                                                   
    \end{tikzpicture}
\end{minipage}
\end{center}
\smallskip

\emph{\hi{Left picture:} Penrose diagram of $m<0$ Schwarzschild, or supercharged Reissner--Nordstr\"{o}m  ($e^2>m^2>0$), or fast  Kerr ($a^2>m^2>0$). These solutions have a singularity at $r=0$, but unlike the $m>0$ Kruskal case it is not shielded by an event horizon. Instead, the red lines labeled $\mathcal{H}_C^-$ and $\mathcal{H}_C^+$ are past and future Cauchy horizons with respect to the blue line, indicating a maximal spacelike surface whose initial data give rise to the metrics in question and whose maximal Cauchy development is the grey area. }
\smallskip

\emph{\hi{Right picture:}  Penrose diagram of subcritical Reissner--Nordstr\"{o}m  ($0<e^2<m^2$), whose event and Cauchy horizons (despite the different structure of the singularity) also resemble those of slowly rotating Kerr ($0<a^2<m^2$). The  maximal Cauchy development of the pertinent initial data given on the maximal spacelike hypersurface represented by the blue line labeled $\Sg$ is again colored in grey. It contains past and future event horizons labeled  $\mathcal{H}_E^-$ and $\mathcal{H}_E^+$, drawn in green, but unlike  the $m>0$ Schwarzschild case  the singularity they are supposed to shield cannot be reached directly from the  maximal Cauchy development, which is bounded by the various fictitious boundaries $\CJ^{\pm}$, $i^{\pm}$, and $i^0$, which lie at infinity, as well as by the Cauchy horizons  $\mathcal{H}_C^{\pm}$, drawn in red, which can be reached in finite proper time.\footnote{ 
This diagram can be infinitely extended  in both directions (Hawking \& Ellis, 1973, pp.\ 158, 165): to the north, another grey area folds inside the upper two red line segments, and similarly to the south, \emph{et cetera}, but we do \emph{not} do so here.}}

\noindent  Despite the different space-times they apply to, the outcomes of the Penrosian version and the \pde\ version of both weak and strong cosmic censorship are once again the same, \emph{mutatis mutandis}:\footnote{For $m<0$ Kruskal the initial data are not complete in this case, so strictly speaking the cosmic censorship conjectures do not apply here and their falsity is unimportant. Nonetheless, they can be stated and the comparison is instructive.} 
\begin{itemize}
\item  $m<0$ Kruskal (etc.): For the Penrosian total space-time the difference between weak and strong  cosmic censorship  fades since
 $J\inv(\CI^+)=M\cup\CI^+$, which, like $M$ itself is not globally hyperbolic: wherever one tries to place a partial Cauchy surface $\Sg$ (such as the blue line), above the surface  inextendible causal curves can be drawn that enter $i^+$ or $\CI^+$ in the future and enter the singularity at $r=0$ in the past, without  crossing $\Sg$. Similarly, below $\Sg$ one may draw  inextendible causal curves converging to the
 singularity in the future, and to  $i^-$ or $\CI^-$ in the past, which once again do not cross $\Sg$. Thus neither weak nor strong cosmic censorship  holds for this space-time.
 
The \pde\ picture applies to the grey area, which is the \mghd\ of the initial data given on the blue line marked $\Sg$  in the left-hand Penrose diagram. Then
 weak cosmic censorship fails because future null infinity $\CI^+$ is clearly incomplete: null geodesics terminate at the Cauchy horizon (where they ``fall off' space-time) and hence are incomplete. On the other hand, strong cosmic censorship fails because the grey space-time, though globally hyperbolic (in contrast with the entire space as we have just seen), is evidently (smoothly--even analytically) extendible, namely by the total space.  Though they do not coincide, we see  that strong and weak cosmic censorship are closely related: future incompleteness of null geodesics at null infinity 
happens because the \mghd\ is extendible.
 
 \item  Subcritical Reissner--Nordstr\"{o}m  ($0<e^2<m^2$):  for both Penrose and the \pde\ people strong cosmic censorship fails whereas the weak version holds.
  In the Penrosian version the total space  fails to be globally hyperbolic because of the part above the grey area (i.e.\ beyond the future Cauchy horizon $\mathcal{H}_C^+$): one has past-directed inextendible causal curves that (backwards in time) end up in the singularity and hence never cross $\Sg$ (e.g.\ those crossing the upper left, NW-pointing red line from N to SW). 
  Morally, weak  cosmic censorship  holds because of the future event horizon $\mathcal{H}_E^+$, which shields the upper $r=0$ singularity above it, 
  but legally this is only the case if we stop the Penrose diagram at the past Cauchy horizon $\mathcal{H}_C^-$, as we have  done in drawing the picture (for otherwise causal curves below it may crash at the  lower $r=0$ singularity and hence never reach $\Sg$). 
  
  The \pde\ view is cleaner here: roughly speaking, as in the $m>0$ Kruskal or Schwarzschild case (but unlike the $m<0$ case) future null infinity $\CI^+$ ends at future timelike infinity $i^+$ and hence is complete, so that weak  cosmic censorship holds. Strong  cosmic censorship, on the other hand, fails because the \mghd\ (marked in grey) is clearly extendible (namely into the Penrosian space-time!). 
\end{itemize}
More generally, if the strong Penrosian conjecture fails for some space-time $(M_P,g_P)$, then its lack of global hyperbolicity  typically occurs because $(M_P,g_P)$ is an extension of the \mghd\ $(M,g)$ of some given initial data, whose Cauchy surface $\Sg$ fails to be one for  $(M_P,g_P)$.
Similarly, if $J\inv(\CI^+)$ is not globally hyperbolic (so that there is a naked singularity), $M_P$ usually comes from extending some $(M,g)$, as above, whose Cauchy surface becomes a partial  Cauchy surface in $M_P$, with an associated future Cauchy horizon that cuts off $\CI^+\cap\til{M}$, causing its incompleteness.\footnote{However, these aren't rigorous deductions: there are pathological cases where strong cosmic censorship holds whilst the weak version fails. See the Penrose diagram at the end of \S 2.6.2 of Dafermos \& Rodnianski (2008) for an example.} 
As already mentioned, any counterexamples are believed to be ``non-generic'', assuming of course that the conjectures hold! 

Such reasoning, which applies to many case studies,  also suggests a compromise between the Penrosian and \pde\ versions of cosmic censorhsip: informally  one might say that, in physically reasonable space-times, weak cosmic censorship postulates the \emph{appearance and stability of event horizons}, whereas  strong cosmic censorship requires the \emph{instability and ensuing disappearance of Cauchy horizons}.
\section{Epilogue: Penrose's final state conjecture}\label{FSC}
In practice, the cosmic censorship conjectures are not put in the full  generality of either Penrose's own version as expressed by Theorem \ref{P79theorem} or of the \pde\ version as Definition \ref{CSfed1}, 
but are posed in the context of black holes as they are expected to occur in the universe, i.e.\ as described by something like the Kerr metric (at least outside its Cauchy horizon and more safely even outside its event horizon). As such, on the one hand
 they gain focus, but on the other hand they can be thought of as
 forming part of a broader  conjecture that also originated with Penrose himself and is often called the \emph{final state conjecture}:\footnote{This is sometimes stated somewhat differently, in that `generic asymptotically  flat vacuum initial data (\ldots) evolve to a solution which either disperses (in which case there are no black holes) or else eventually asymptotes to
  finitely many Kerr solutions (...) moving away from each other' (Coley, 2019, p.\ 78). See also the fascinating lecture by Klainerman (2014).
  }
 \begin{quote}
\begin{small}A body, or collection of bodies, collapses down to a size comparable to its Schwarzschild radius, after which a trapped surface can be found in the region surrounding the matter. Some way outside the trapped surface region is a surface which will ultimately be the absolute event horizon. But at present, this surface is still expanding somewhat. Its exact location is a complicated affair and it depends on how much more matter (or radiation) ultimately falls in. We assume only a finite amount falls in and that {\sc gic} is true. Then the expansion of the absolute event horizon gradually slows down to stationarity. Ultimately the field settles down to becoming a Kerr solution (in
the vacuum case) or a Kerr-Newman solution (if a nonzero net charge is trapped in the ``black hole''). (Penrose, 1969, pp.\ 1157--1158)
\end{small}
\end{quote}
Here {\sc gic} refers to what Penrose (1969) called the \emph{Generalized Israel Conjecture},\footnote{The reference is to Israel (1967, 1968). See Israel (1987), \S 7.9 and Thorne (1994), Chapter 7, for interesting history.}
which states that:
 \begin{quote}
\begin{small}
 if an absolute event horizon develops in an asymptotically flat space-time, then the solution exterior to this horizon approaches a Kerr-Newman solution asymptotically with time. (Penrose, 1969, pp.\ 1156)
\end{small}
\end{quote}
 In the stationary case, which is what Israel himself conjectured and proved under fairly restrictive assumptions, this would simply say that the solution exterior to this horizon \emph{equals} a Kerr-Newman solution. As such, the conjecture  is an outgrowth of what (following Wheeler) used to be called the ``no hair'' property of black holes, to the effect that stationary black holes (and eventually all black holes)
  are characterized by by just three parameters, viz.\ mass, angular momentum, and electric charge. As such, the final state conjecture incorporates not only weak cosmic censorship (notably in the compromise version suggested at the end of the previous section) but also what in the \pde\ literature is called \emph{Kerr stability},  as well as \emph{Kerr rigidity}. The former is
   the conjecture that generic  perturbations of the  initial data
 for the  Kerr metric lead to a \mghd\ that is close to the original one (at least outside the event horizon).\footnote{There is certain  numerical evidence for this (Zilh\~{a}o et al., 2014), but mathematical results so far are  preliminary (Dafermos,  Holzegel, \&  Rodnianski, 2019b; 
Giorgi, Klainerman, \& Szeftel, 2020), except for positive cosmological constant and small $a$  (Hintz \& Vasy, 2018), where the problem is solved. 
Even the Schwarzschild case is still open, despite impressive progress (Klainerman \& Szeftel, 2017; Dafermos,  Holzegel, \&  Rodnianski, 2019a).
} This would generalize the remarkable theorem on the stability of Minkowski space-time (Christodoulou \& Klainerman, 1993), which launched the modern era in \pde-oriented mathematical relativity. 
The latter  is a more modest version of the no-hair or black hole uniqueness theorems of Israel, Carter, Hawking, Robinson, and others,\footnote{These theorems are reviewed in Hawking \& Ellis (1973), \S 9.3, Carter (1979, 1986), Heusler (1996), Robinson (2009), Chru\'{s}ciel,  Lopes Costa, \& Heusler (2012), and Cederbaum (2019).  See also Cardoso \&  Gualtieri (2016) for possible  tests.
} where short of proving the stationary case of the above {\sc gic}, which requires unphysical analyticity assumptions, one tries to show that at least stationary solutions to Einstein's vacuum (electrovac) equations that are close to Kerr (--Newman) actually coincide with the latter.\footnote{See e.g.\
Alexakis,  Ionescu, \& Klainerman (2014) as well as the review by Ionescu \&  Klainerman (2015).}
\smallskip

In conclusion, ``the discovery that black hole formation is a robust prediction of the general theory of relativity'' still lies in the future as far as mathematical proof is concerned. Penrose's Nobel Prize was effectively awarded for a theorem and a conjecture, but it was fully deserved in every conceivable way!
\appendix
\section{A potted early history of ``black hole'' (by Erik Curiel)} 
In reply to a query in an early  draft of this paper concerning  the origin of the definitions $\mathcal{B}:= M\backslash J^-(\CI^+)$ of a black hole (region) and of the event horizon as its boundary $\mathcal{H}_E^+:=\p \mathcal{B}$, see 
\er{BHRWHR} and \er{defEHplus}, phrased as: `Such definitions are routinely used in e.g.\ Carter (1971a) and Hawking \& Ellis (1973), \S 9.2, but who stated them first?',  Erik Curiel very kindly supplied the following information.\footnote{As in the rest of the paper, single quotation marks below denote literal quotation whereas double ones are scare quotes.}  As we see, the question still remains somewhat open. A puzzling point is that
although Penrose himself would have been the obvious person to state these definitions mathematically, apparently he did not do so!
\begin{quote}
\begin{small}
Penrose (1968) defines (p.\ 188), an event horizon as the boundary of the chronological past of
a timelike curve (essentially the same definition, including the name, as given by Rindler 1956),
and notes (p.\ 206) that $r = 2M$ in Schwarzschild is one. The term ``black hole'' does not appear
in that essay, nor any definition remotely like `the complement of the causal past of future null
infinity'. Given the magisterial depth and encyclopedic scope of that essay, I must conclude that
the definition was not then yet extant. Penrose (1969) does use the term ``black hole'' (the first use
of it I know in the general relativity literature, though it reportedly was used in the early 1960s by
Dicke in discussion with a popular science writer), but he always encloses it in scare quotes, leading
me to believe that the name and the general concept both were still inchoate. This is buttressed
by the fact that he does give here (p.\ 1146, footnote 3) the classic definition of an `absolute event
horizon' (the boundary of the chronological past of future null infinity), but he does not explicitly
link it to the term ``black hole''. That is the first appearance of the classic definition I know of in
the literature.

Ruffini and Wheeler gave a series of lectures in September 1969 at the \emph{Interlaken Colloquium on
the Significance of Space Research for Fundamental Physics} (Interlaken, Switzerland), one of which
was entitled `Black Holes', at least according to the expanded version of the lectures published as
Ruffini and Wheeler (1971a), from which Ruffini and Wheeler (1971b) was adapted. This is the first
use of the term I have been able to find recorded in a public forum in the relativity community.
They explain the idea in informal, intuitive terms. Bardeen (1970), received 23 January 1970,
has ``black hole'' in the title, the first publication I know of to do so. He introduces ``black hole''
using scare quotes and equates it with a `collapsed object'. Israel (1971), originally read at 
 the \emph{Gwatt Seminar on the Bearings of
Topology upon General Relativity} on 19 May 1970, uses ``black hole'' without defining it, but it is clear from context that he
means something like `system that quickly settles down so that its exterior is modeled by the Kerr
solution'.  Christodoulou (1970), received 17 September 1970, uses the term without blushing,
not even an informal gloss given for its meaning. He simply begins by talking of a `black hole
[with] angular momentum' without even citing Kerr (1963). Three papers then appear in 1971
with ``black hole'' in the title, Penrose and Floyd (1971), received 16 December 1970, Carter (1971),
received 18 December 1970, and Hawking (1971), received 11 March 1971. They are the only other
papers from 1971 I can find related to the topic whose authors plausibly could have proposed
the classic definition. Penrose and Floyd (1971) uses scare-quotes around the first use of ``black
hole''; their discussion relies only on the event horizon and ergosphere (which they refer to as the
`stationary limit') defined by the Kerr metric, with no attempt at (or mention of the possibility
of) generalization. Carter (1971) does not give a formal definition of ``black hole'', but he does give
an informal definition of `domain of outer communication', and says (p.\ 331) that `\! ``black holes''
[are] regions of space-time beyond the domain of outer communication.' Note the scare-quotes.
Hawking (1971) uses scare-quotes as well, going out of his way to assimilate the idea of a black
hole to one more widely known (`there are initially two collapsed objects or ``black holes''', p.\ 1345).
He also comes achingly close to defining a black hole as a connected component of the complement
of the causal past of future null infinity, but never quite does it. He rather says things like, 
`On $\Sg_i$ [a spacelike hypersurface], there will be two separate regions, $B_1$ and $B_2$ which contain closed,
trapped surfaces (\ldots)  Just outside $B_1$ and $B_2$ will be two two-spheres which are the intersection
of $\dot{J}^-(\CI^+)$ with $\Sg_i$.' The first explicit definition I know of ``black hole'' as `connected component of the complement of the causal past of future null infinity' is in Hawking (1972). But I am not
confident that I have found all relevant sources in the literature; even if I have, one cannot be
confident based only on this that Hawking was the one who finally put all the pieces together.
 
(Erik Curiel, private communication, January 6, 2021, reprinted with permission)
\end{small}
\end{quote}
\newpage

\noindent \textbf{Acknowledgement.} This paper grew out of a seminar in honour of Roger Penrose and his Nobel Prize at Nijmegen on October 28th, 2020, at the kind invitation of Timothy Budd. The author is grateful to Carlo Rovelli for encouragement to publish it in \emph{Foundations of Physics}, as well as to B\'{e}atrice Bonga, Jeremy Butterfield, Erik Curiel, Juliusz Doboszewski, John Earman,  Dejan Gajic, and Dennis Lemkuhl for helpful feedback on earlier drafts. I also learned a lot form various seminars by Martin Lesourd.
\smallskip

\addcontentsline{toc}{section}{References}
\begin{small}

\end{small}
\end{document}